\documentclass{aa}  

\usepackage{graphicx}
\usepackage{txfonts}
\usepackage{subcaption}
\usepackage[normalem]{ulem}
\usepackage{xcolor}
\usepackage{multirow}
\usepackage{caption}
\usepackage[colorlinks=true, linkcolor=blue, citecolor=blue, urlcolor=blue]{hyperref}
\usepackage{array} % 提供 \extrarowheight
\renewcommand{\arraystretch}{1.2}
\begin{document}

   \title{Metal enrichment of galaxies \\ in a massive node  of the Cosmic Web at $z \sim 3$}
    \titlerunning{Metallicity in cosmic web node}
   % \subtitle{I. Overviewing the $\kappa$-mechanism}

   \author{Xiaohan Wang
          \inst{1, 2}, 
          S. Cantalupo\inst{2}, % \fnmsep\thanks{Just to show the usage of the elements in the author field}
          Weichen Wang\inst{2},
          M. Galbiati\inst{2},
          Charles C. Steidel \inst{4},
          A. Pensabene\inst{2},
          Shude Mao\inst{3},
          A. Travascio \inst{2},
          T. Lazeyras \inst{2},
          N. Ledos \inst{2},
          G. Quadri \inst{2}
          }
    \authorrunning{Xiaohan Wang et al}

   \institute{Department of Astronomy, Tsinghua University, Beijing     100084, China
         \and
              Dipartimento di Fisica, Università degli Studi di Milano-Bicocca, Piazza della Scienza 3, 20126 Milano, Italy
         \and 
             Department of Astronomy, Westlake University, Hangzhou 310030, Zhejiang Province, China
        \and
            Cahill Center for Astronomy and Astrophysics, California Institute of Technology, MS 249-17, Pasadena, CA 91125, USA
             }

   \date{Nov. 24, 2025}

% \abstract{}{}{}{}{} 
% 5 {} token are mandatory
 
  \abstract
  % % context heading (optional)
  % % {} leave it empty if necessary  
  %  {}
  % % aims heading (mandatory)
  %  {}
  % % methods heading (mandatory)
  %  {}
  % % results heading (mandatory)
  %  {}
  %  % conclusions heading (optional), leave it empty if necessary 
   {We present the mass-metallicity relation for star-forming galaxies in the MUSE Quasar Nebula 01 (MQN01) field, a massive cosmic web node at $z \sim 3.245$,
   hosting one of the largest overdensities of galaxies and AGNs found so far at $z > 3$.
   Through James Webb Space Telescope (JWST) Near Infrared Spectrograph (NIRSpec) spectra and images from JWST and Hubble Space Telescope (HST), we identify a sample of 9 star-forming galaxies in the MQN01 field with detection of nebular emission lines ($\rm H\beta$, [OIII], $\rm H\alpha$, [NII]), covering the mass range of $\rm 10^{7.5}M_\odot - 10^{10.5}M_\odot$.
   % with a median offset of $\sim 0.36$ dex off the referred star formation main sequence. 
   We present the relations of the emission-line flux ratios versus stellar mass for the sample and derive the gas-phase metallicity based on the strong line diagnostics of [OIII]$\lambda5008$/$\rm H\beta$ and [NII]$\lambda6585$/$\rm H\alpha$.
   Compared to the typical, field galaxies at similar redshifts, MQN01 galaxies show relatively higher [NII]$\lambda6585$/$\rm H\alpha$ and lower [OIII]$\lambda5008$/$\rm H\beta$ at the same stellar mass, %and enriched metallicity 
   which implies a higher metallicity by
    about $0.25\pm 0.07$ dex 
   with respect to the field mass-metallicity relation.
   These differences are decreased considering the ``Fundamental Metallicity Relation'', i.e. if the galaxies' Star Formation Rates (SFR) are also taken into account.
   %The fundamental metallicity relation of the MQN01 is more consistent with field galaxies. %  with a mean deficiency of $\sim 0.07$ dex.
   %We suggest that the distributions shown on the mass-metallicity relation and the fundamental metallicity relation 
   We argue that these results
   are consistent with a scenario in which galaxies in overdense regions assemble their stellar mass more efficiently (or, equivalently, start forming at earlier epochs) compared to field galaxies at similar redshifts.}


\maketitle
%
%-------------------------------------------------------------------
\section{Introduction}
\label{intro}
Galaxies evolve within ecosystems, exchanging material with the environments through processes including gas accretion, mergers, galaxy interactions, and feedback from stars and active galactic nuclei (AGN). These processes vary across environments and play critical roles in shaping galaxy properties \citep{Dekel2009, Somerville2015, Tumlinson2017}. At $z < 1$, observations have shown that galaxies living in dense environments such as clusters tend to be more massive with lower star formation rates \citep[e.g.][]{Baldry2006, Peng2010, vanderBurg2020, McNab2021}.
The relative importance of different physical mechanisms in shaping galaxy properties is still unclear. At higher redshifts, the cosmic environment is different from the local universe, with much higher molecular and neutral gas density, more intense star formation activity and potentially stronger feedback \citep[e.g.][]{Shapley2003, Steidel2010, Tacconi2010, Behroozi2013, Lilly2013, Madau2014, Popping2014, Speagle2014, Genzel2015}. To understand the properties of today's most massive galaxies, it is crucial to study their progenitors under these distinct environmental conditions.
The progenitors of present-day massive galaxies living in clusters are found in environments called ``protoclusters'', which work as natural laboratories to investigate the environmental effects.
Observations of protoclusters have shown that galaxies tend to have higher star formation rates with larger fractions of massive galaxies and active galactic nuclei, consistent with stronger interactions and gas accretion that may contribute to more intense star formation activities \citep[e.g.][]{Steidel2005, K2013a, K2013b}.

Metallicity is a key property to understand galaxy star formation histories. The metal enrichment and dilution are highly affected by galaxy star formation, feedback, gas inflows and galaxy interactions. Metallicity has been found to tightly relate to galaxy stellar mass, known as the mass-metallicity relation, across a wide mass range \citep[e.g.][]{Tremonti2004, Erb2006, Maiolio2008, Steidel2014, Sanders2021, Henry2021, Li2023, Langeroodi2023, Stephenson2024, Pallottini2025, Li2025}.
Even though metallicity is found to decrease with redshifts, the relation exists across a wide redshift range (up to $z > 6$) with potentially different slopes \citep[e.g.][]{Sanders2021, Nakajima2023, Curti2024, Sarkar2025, Li2025}.
How metallicities differ between galaxies in protoclusters or in the fields, however, remains poorly understood. Observations have reported inconsistent conclusions at similar redshifts, and the difference appears to evolve with redshift. 
Observations of the X-ray cluster XCS2215 ($z = 1.46$) show metallicity enhancement \citep{Adachi2025}. Some observations for $z \sim 2$ protoclusters, such as BOSS1244, report metallicity deficit at the high mass end and enhancement at the low mass end \citep{Kulas2013, Kacprzak2015, Chartab2021, Sattari2021, Wang2022}. They suggest that these are consistent with a combination of metal enrichment through feedback and dilution by cold gas accretion. Meanwhile, observations of the SpiderWeb protocluster report a metallicity enhancement \citep{Shimakawa2015, P2023}.
At higher redshifts ($ 5 < z < 7$), a recent work shows enhanced metallicity of $\sim 0.2$ dex on the mass-metallicity relation \citep{Li2025}.
% . However, when analyzing using fundamental metallicity relation (FZR), which adopts a combined parameter including stellar mass and star formation rate, the same dataset shows a $\sim$ 0.2 dex deficit in metallicity \citep{Li2025}, suggesting more complex interplays among the physical processes in chemical evolution.
Despite these growing efforts, metallicity observations for protoclusters are still limited across a large range of cosmic time, stellar masses and galaxy overdensities.

In this work, we present the mass-metallicity relation and emission line properties obtained in a field containing the largest overdensity of galaxies and AGNs found so far at $z \sim 3$, 
the MUSE Quasar Nebular 01 field (MQN01). MQN01 is a large scale galaxy overdensity around the $z \sim 3.25$ QSO CTS G18.01, first detected as filamentary-shaped $\rm Ly\alpha$ emission structure \citep{Borisova2016} and later confirmed by deeper observations (Cantalupo et al., in prep.).
MQN01 has been observed by multi-wavelength surveys and has shown high galaxy overdensity and a large fraction of massive galaxies of AGNs with abundant molecular gas \citep{Pensabene2024, Galbiati2025, Travascio2025}. MQN01 also contains the largest disk galaxy found so far at $z > 3$, called ``Big Wheel'', with an optical diameter larger than 30 kpc \citep{Wang2025}.
In this work, we present a study of galaxies detected in the same field using JWST NIRSpec observations, and analyze their emission line properties.
% With follow-up observations from JWST NIRSpec, we get slit spectra for galaxies in MQN01 do the analysis for emission line properties.
In Sec. \ref{data_method} we introduce the observations of MQN01, sample identification, and measurements of galaxy properties.
% In Sec \ref{obs} we introduce the imaging and spectra observations of MQN01 and the sample. In Sec \ref{method} we introduce the methodology of emission line analysis, measurements of stellar mass, star formation rate and gas-phase metallicity. 
In Sec. \ref{result} we present our main results, i.e. the relations between line flux ratios and metallicity versus stellar mass, with comparisons with field galaxies at similar redshifts. In Sec. \ref{dis} we discuss possible mechanisms related to environmental effects on metal enrichments, and present the fundamental metallicity relation. 
% present the fundamental metallicity relation and discuss observational effects(?) and possible mechanisms that result in the differences.
We summarize in Sec. \ref{conc}. We adopt the AB magnitude system \citep{Oke1983}.
We adopt a Flat$\Lambda$CDM cosmology with $H_0 = 67.66\rm \ km\ s^{-1}\ Mpc^{-1}$, $\Omega_m = 0.31$ and $\Omega_{\Lambda} = 0.69$ \citep[Planck 2018, ][]{Planck2020} \footnote{All cosmological calculations in this work are performed using the \texttt{Planck18} cosmology implemented in the \texttt{astropy.cosmology} package \citep{Astropy2022}.}.

%--------------------------------------------------------------------
\section{Data sample and Methods}
\label{data_method}
\subsection{JWST observation of MQN01 protocluster}
\label{obs}

    \begin{figure*}
   \centering
   \includegraphics[width = 1.8\columnwidth]{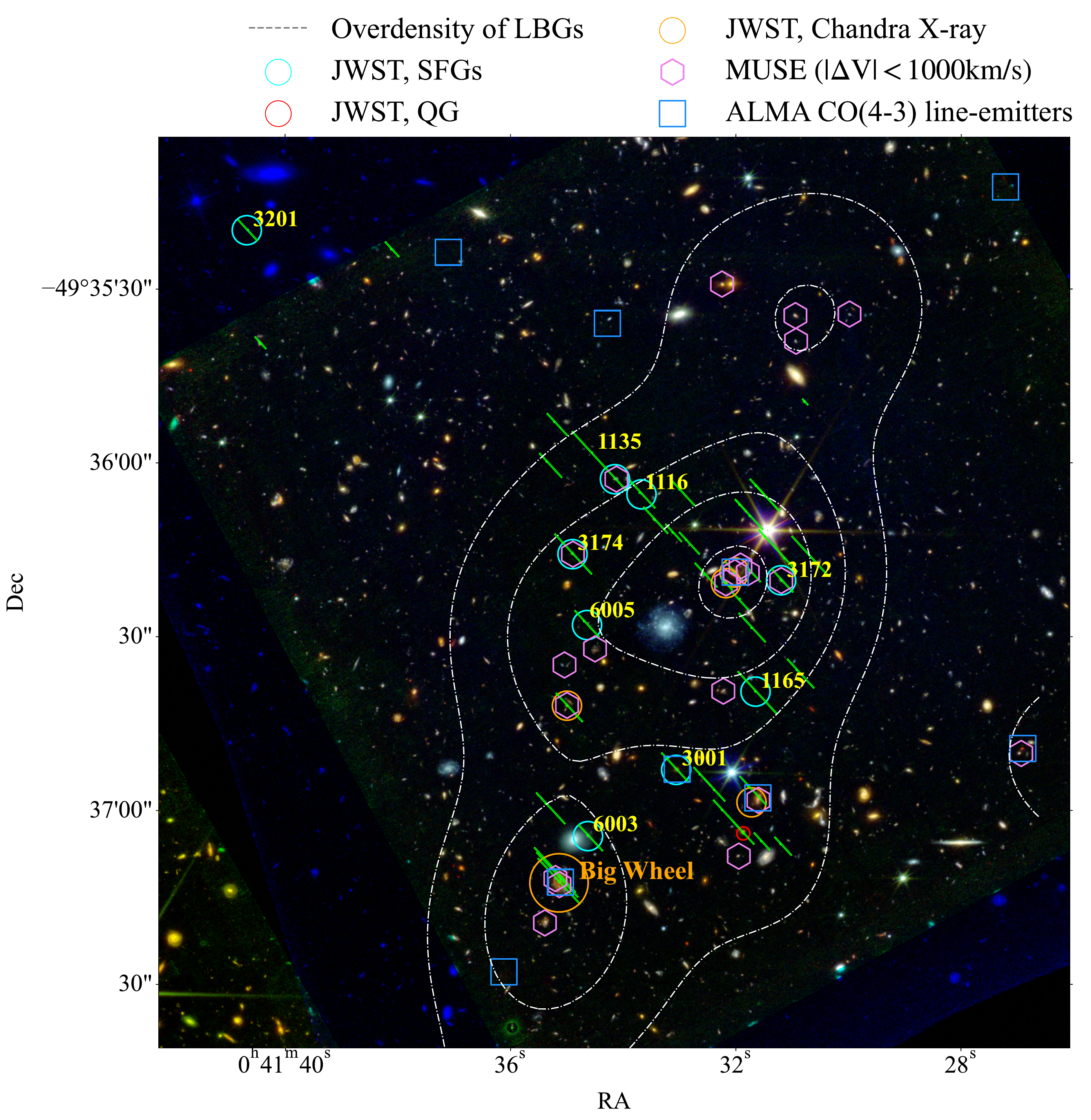}
   \caption{Slit design of the JWST Program GO 1835 (PI: Cantalupo) targeting the MQN01 protocluster field (see text for details on slit location selection).
   %The locations of the slits have been selected based on position of bright Ly$\alpha$ diffuse and extended emission detected by deep MUSE observations (Cantalupo et al, in prep.). In some cases, slits have been located on bright continuum sources without prior knowledge of their redshift. 
   The composite false-color image of the field is created with HST F814W (0.8$\rm \mu m$; blue), JWST F150W2 (1.5$\rm \mu m$; green), and JWST F322W2 (3.2$\rm \mu m$; red). The dashed contours show the overdensity of galaxies \citep{Galbiati2025}. The distribution of samples identified from MUSE observations \citep{Galbiati2025}, Chandra X-ray observations \citep{Travascio2025} and ALMA \citep{Pensabene2024} are also included.
    The sample of identified star-forming galaxies detected with emission lines is marked with cyan circles. Galaxies with X-ray emission are marked in orange circles. A quiescent galaxy serendipitously found with the JWST NIRSpec observations (Wang et al., in prep.) is marked with a red circle.}
    \label{sky_region}%
    \end{figure*}
% The color density shows the distribution of $\rm Ly\alpha$ emission.

% 3001', '1135', '3174', '3172', '3201', '6003', '1165', '1116', '1177', '6005'

\begin{figure*}
  \centering
  % 自定义两行的目标高度（按需调整）
  \newlength{\rowoneht}\setlength{\rowoneht}{0.22\textheight}
  \newlength{\rowtwoht}\setlength{\rowtwoht}{0.22\textheight}
  % ==== 第一行：单图，跨全宽 ====
  \includegraphics[width=\linewidth]{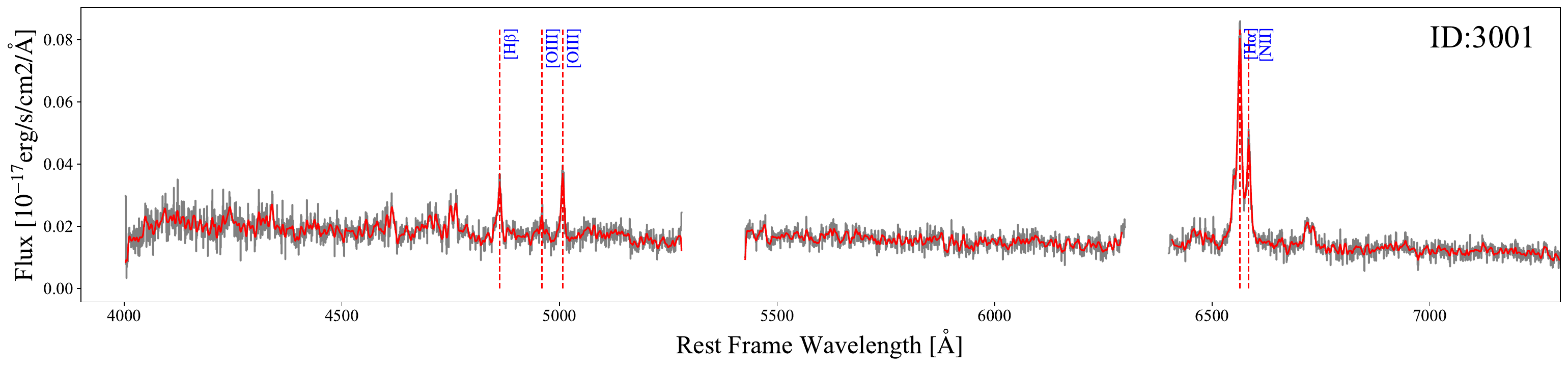}
  % ==== 第二行：两图并排、等高、零间隙 ====
  % 用两个 minipage，行末加 % 去掉空白，且不插入 \hfill
  \noindent
   \vspace{10pt}
  \begin{minipage}{0.81\linewidth}
    \includegraphics[height=\rowtwoht, width=\linewidth, keepaspectratio]{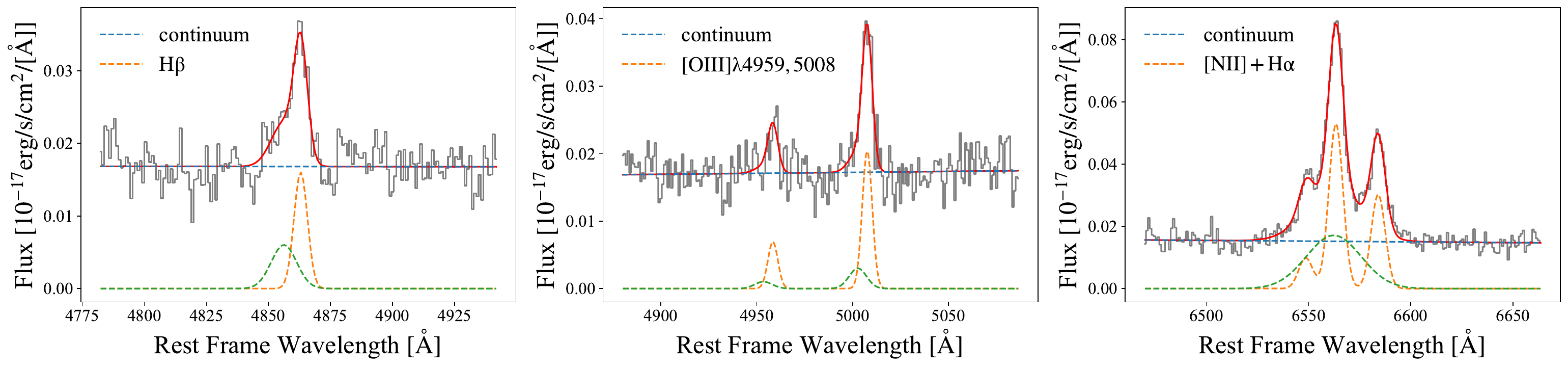}
  \end{minipage}%
  \begin{minipage}{0.19\linewidth}
    \includegraphics[height=\rowtwoht, width=\linewidth, keepaspectratio]{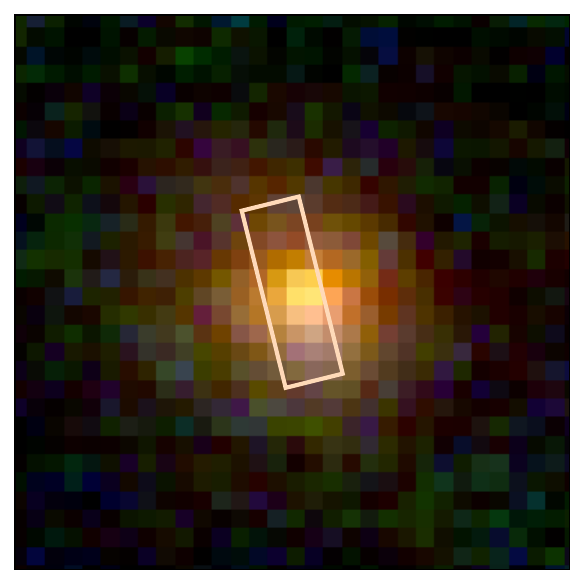}
  \end{minipage}
  % ==== 第一行：单图，跨全宽 ====
  \includegraphics[width=\linewidth]{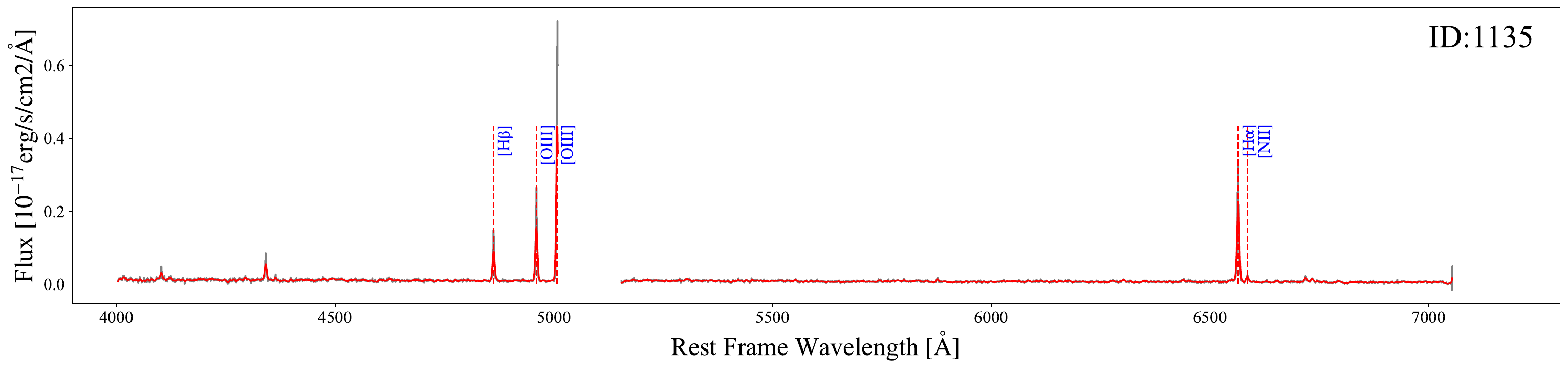}
  % ==== 第二行：两图并排、等高、零间隙 ====
  % 用两个 minipage，行末加 % 去掉空白，且不插入 \hfill
  \noindent
  \begin{minipage}{0.81\linewidth}
    \includegraphics[height=\rowtwoht, width=\linewidth, keepaspectratio]{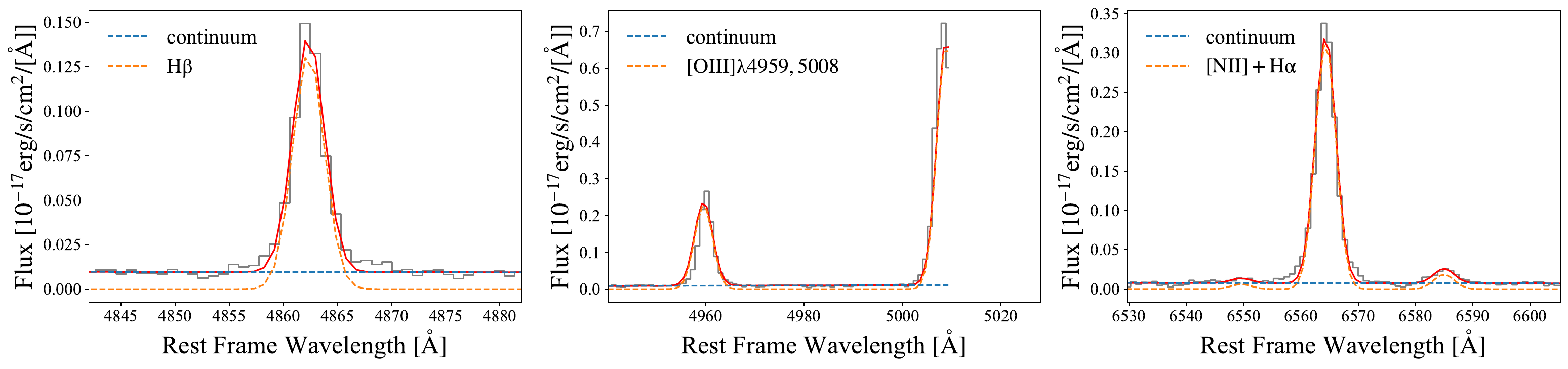}
  \end{minipage}%
  \begin{minipage}{0.19\linewidth}
    \includegraphics[height=\rowtwoht, width=\linewidth, keepaspectratio]{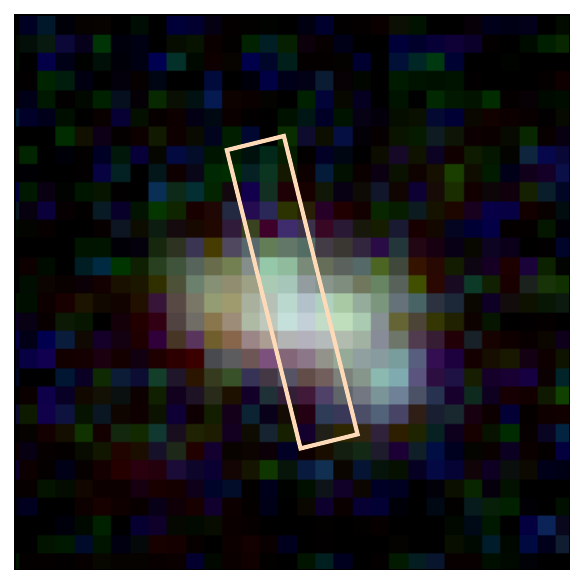}
  \end{minipage}
  \caption{Integrated 1D spectra and composite false-color images of two galaxies in the sample, with slit IDs 3001 and 1135. The flux density is in unit of $10^{-17} \rm erg/s/cm^2/$\AA, and the wavelength is in rest-frame Angstroms. The slit regions where the 1D spectra are extracted are overplotted on the galaxy images as light pink boxes.
  Emision lines are modeled with single Gaussians.
  For galaxy 3001, which shows road emission-line features, an additional Gaussian component is included (shown by the green dotted line). Flux of the narrow Gaussian components are adopted in the analysis for the galaxy 3001.
    The filters used to create the color images are HST F814W (0.8$\rm \mu m$; blue), JWST F150W2 (1.5$\rm \mu m$; green), and JWST F322W2 (3.2$\rm \mu m$; red).
    Spectra and images for the remaining galaxies are given in Fig. \ref{spec_image_full}.}
  \label{spec_image}
\end{figure*}

Previous multi-wavelength observations have provided a multi-wavelength view of galaxies of MQN01, highlighting MQN01 as one of the most overdense protoclusters with high gas density and AGN fractions \citep{Pensabene2024, Galbiati2025, Travascio2025}. The multi-wavelength observations altogether have identified a sample of 26 galaxies, where 21 galaxies are MUSE continuum-selected absorption line galaxies and 5 galaxies are newly uncovered by ALMA detection. 5 of the 26 galaxies are confirmed to have X-ray emission by Chandra observations \citep{Travascio2025}. 

The imaging and spectroscopic data analyzed in this work of the MQN01 field were obtained from HST Program GO 17065 (PI: Cantalupo), VLT/HAWK-I program ID 110.23ZX, (PI: Cantalupo) and JWST Program GO 1835 (PI: Cantalupo). Photometry includes JWST NIRCam filters F150W2 and F322W2 \citep{Rigby2023, Rieke2023}, HST filters ACS/WFC F625W and ACS/WFC F814W \citep{Ford1998}, and HAWK-I filters CH4, H and $\rm K_{s}$ \citep{P2004, Casali2006, K2008, S2011}.
Spectroscopic observations, about 8 hours on average, were conducted using the Micro-Shutter Assembly (MSA) observing mode of JWST NIRSpec, with the F170LP/G235H filter and grating pair, covering wavelength from  $1.66$ to $3.05 \rm \mu m$ at a spectral resolution of $R \sim 2700$. At $z \sim 3.245$, the wavelength coverage corresponds to $3910 - 7185$\AA\, in rest-frame, including key Balmer lines ($\rm H\alpha$
 to $\rm H\gamma$) and metal emission lines ([OIII], [NII], e.g.) \citep{Jakobsen2022, Boker2023}.
The slits were arranged to cover regions of extended $\rm Ly\alpha$ emission \citep[Cantalupo in prep.]{Borisova2016} on the sky, independent of the previous knowledge of associated galaxies at the redshift of the MQN01 structure. In a very limited number of cases and for location for which extended emission was not present, slits have been positioned on known protocluster members, independently on their stellar mass or SFR. Finally, in some cases, ``filler'' slits have been located on continuum sources without knowledge of their redshift. 
% Slit locations were chosen based on the surface density distribution of $\rm Ly\alpha$ emission. 
The slit design on the MQN01 field is shown in Fig. \ref{sky_region}.
The spectra for each slit were reduced and combined using the official jwst pipeline (v1.11.3) with calibration reference file version jwst 1097.pmap.
For each slit, background is modeled by averaging spectra from regions outside of the target galaxy. Spectra near the slit edges were excluded due to noise.
The resulting 1D spectra with identified emission lines and JWST F322W images are presented in Fig. \ref{spec_image} and Fig. \ref{spec_image_full}. Galaxy redshifts are inferred from the observed wavelengths of $\rm H\alpha$ and [OIII]$\lambda5008$ lines.

The new JWST/NIRSpec observations complement the existing multi-wavelength data of MQN01, providing rest-frame optical spectra with prominent emission lines and thereby extending the available galaxy sample. The construction and properties of the final sample are presented in Sec. \ref{sample_selection}.

\subsection{Measurements}
\label{method}
\subsubsection{Emission line flux}

The 1D spectra of galaxies are extracted from the spatial regions identified in the following way.
We determine the galaxy boundary on the slit first by a single Gaussian fitting to the JWST F322W images, and then refine it with $\rm H\alpha$ emission detection. Spectra are integrated within identified galaxy regions.

Emission lines are modeled with single Gaussian components, with flux ratios fixed at 3.06 for [NII] doublets and 2.94 for [OIII] doublets. The kinematic properties, velocity and dispersion, are assumed to be the same for [OIII] doublets, or $\rm H\alpha$ + [NII] doublets. The stellar continuum, which is substantially weaker than the emission lines for our sample, is removed by a linear fitting of the continuum region around the emission lines of interest.
For one galaxy which shows broad emission-line features (ID: 3001, shown in the top panel of Fig. \ref{spec_image}), the emission lines are modeled with two Gaussian components (one extra component for $\rm H\alpha$) + [NII] doublets). Flux of the narrow Gaussian components are adopted in the analysis for the galaxy 3001.

\subsubsection{Stellar mass}
\label{mass_cal}
We derive stellar masses through SED fitting with CIGALE v2025.0 \citep{Burgarella2005, Noll2009, Boquien2019}, using photometry from HST F814W, F625W, and F160W, HAWK-I $\rm K_{s}$, and JWST F150W2 and F322W2.
For one galaxy that is not observed by JWST/NIRCam (ID: 3201), we use photometry from HST F814W, HAWK-I H, CH4 and $\rm K_{s}$.
A delayed-$\tau$ star formation history with a large fixed value of $\tau=100~\mathrm{Gyr}$ (effectively constant SFR) is assumed, together with a \citet{Chabrier2003} initial mass function, \citet{Bruzual2003} stellar population synthesis models, and the \citet{Calzetti2000} dust attenuation law. Both stellar and nebular metallicities are fixed to solar metallicity (0.02).% ($Z_{\odot}$). 
The input parameters in the SED fitting are summarized in Table \ref{SED_para_table}.
We adopt the median (50th percentile) stellar mass of the Bayesian posterior distribution in log space, and uncertainties by the 16th–84th percentile range.

To address the robustness of stellar mass estimates, we have performed a series of SED fittings with CIGALE, including different star formation histories (with and without bursts), metallicities, ionization parameters ($\log U$ and $n_e$), and grid setups for the star formation history parameters.
The stellar masses remain broadly consistent. In addition, we also carried out runs without nebular emission to provide conservative upper limits on the stellar mass, which show consistent results. % The detailed comparisons and illustrative figures are deferred to Appendix X, while here 
We have also tested a combination of $Z_*= 0.004$\footnote{Referred to ``0.27$Z_\odot$+SMC'' in \citet{Reddy2018, Shapley2023a}, where $Z_\odot = 0.014$.} and the Small Magellanic Cloud (SMC) extinction curve \citep{Gordon2003}, which is suggested for high redshift and low-metallicity galaxies \citep{Reddy2018, Du2018, Shapley2023a}.
The derived stellar masses are relatively lower than the combination of solar metallicity + \citet{Calzetti2000} extinction, while the differences are small and do not affect our main conclusions.
We adopt the masses derived with a configuration similar to that commonly used in the literature, especially for a better consistency with the reference field galaxy datasets \citep{Sanders2021, Li2023}.
Figures utilizing $Z_*= 0.004$ + SMC stellar masses are shown in Appendix \ref{new_figures_SMC} for reference.

\begin{table}[]
    \centering
    \begin{tabular}{{p{3cm} p{5cm}}}
        \hline
        \hline
        Parameter & Value \\
        \hline
        \multicolumn{2}{c}{sfhdelayed} \\
        \hline
         $\tau$ & 10Gyr \\
        Age & 20, 30, 40, 50, 65, 80, 95, 110, 130, 150, 170, 200, 230, 260, 300, 350, 400, 460, 520, 600, 680, 760, 840, 920, 1000, 1100, 1200, 1350, 1500, 1650, 1800, 1900 (Myr)\\
        $f_{\rm burst}$ & 0.0 \\
        \hline
        \multicolumn{2}{c}{bc03} \\
        \hline
        imf & 1 (Chabrier) \\
        metallicity & 0.02 \\
        seperation\_age & 10 Myr\\
        \hline
        \multicolumn{2}{c}{nebular} \\
        \hline
        logU & -3.8, -3.2, -2.6, -2.0, -1.5\\
        zgas & 0.02\\
        ne & 100 \\
        $f_{\rm esc} $ & 0.0 \\
        $f_{\rm dust} $ & 0.0 \\
        lines\_width & 500.0 km/s \\
        \hline
        \multicolumn{2}{c}{dustatt\_calzleit} \\
        \hline
        E\_BVs\_young & 0.00, 0.05, 0.1, 0.15, 0.2, 0.25, 0.30, 0.45\\
        E\_BVs\_old\_factor & 0.44\\
        uv\_bump\_amplitude & 0.0\\
        powerlaw\_slope & 0.0\\
        filters & galex.FUV \& generic.bessell.B \& generic.bessell.V \\
        \hline
        
    \end{tabular}
    \caption{Input parameters for the SED fitting code CIGALE.}
    \label{SED_para_table}
\end{table}

\subsubsection{Star formation rate}
\label{sfr_cal}
The star formation rate is measured from integrated $\rm H\alpha$ flux, corrected for both aperture effects and dust attenuation. We use
\begin{equation}
    \mathrm{SFR[M_\odot yr ^{-1}]} = 10^{-41.35}\times L_{\rm H\alpha} [\rm erg/s]
\end{equation}
assuming a \citet{Chabrier2003} IMF \citep{Hao2011}. The distribution of the sample on the star formation main sequence is presented in Fig. \ref{SFMS}.

It has been shown that galaxies with low metallicities tend to produce more $\rm H\alpha$ emission at the same SFR, with the conversion factor increasing with metallicity \cite{KC2025}. A lower conversion factor may be suggested for high redshift galaxies with lower stellar masses and lower metallicities \cite{Shapley2023a}.
We also calculate SFR using $\mathrm{SFR[M_\odot yr ^{-1}]} = 10^{-41.64}\times L_{\rm H\alpha} [\rm erg/s]$, which is derived from low metallicity BPASS models \citep{Theios2019}, 
To maintain consistency with the reference field galaxy datasets\citep{Sanders2021, Li2023}, we adopted the conversion factor of $10^{-41.35}$ throughout the analysis.
Figures utilizing the lower conversion factor, combined with $Z_*= 0.004$ + SMC stellar masses (see Sec. \ref{mass_cal}), are shown in Appendix \ref{new_figures_SMC} for reference.

To correct for slit loss, we perform rectangular aperture photometry over the slit-covered region of each galaxy and calculate the ratio between the galaxy flux falling within the slit region used for the spectrum extraction and the total galaxy flux.
Aperture correction factors were calculated using F322W2 and F150W2, and are applied separately to $\rm H\alpha$ and $\rm H\beta$ fluxes.
For the galaxy that is not observed by JWST/NIRCam (ID: 3201), we estimate the correction using the HST F814W image.
The central wavelength of F814W is much shorter than the redshifted $\rm H\alpha$ wavelength and may lead to underestimation. However, the image resolution of HST is also worse than JWST, leading to overestimation conversely. We simply adopt this value for aperture correction for this galaxy and note this in Figure captions and following texts. Both $\rm H\alpha$ and $\rm H\beta$ are corrected using the same factor for this galaxy.

The star formation rate is dust-corrected with \citet{Cardelli1989} law assuming an intrinsic Balmer decrement of 2.79.
For galaxies whose $\rm H\beta$ are too faint to be detected, we adopt uncorrected SFR as lower limits.

\subsubsection{Gas-phase metallicity}
Gas-phase metallicity is derived using calibrations in \citet{Bian2018}, with emission line diagnostics, N2 ($\rm \log [NII]\lambda 6585/H\alpha$), O3 ($\rm \log [OIII]\lambda 4959, 5008/H\beta$) and O3N2 ($\rm \log (([OIII]\lambda 5008/H\beta)/([NII]\lambda 6585/H\alpha))$), following the empirical relations
\begin{align}
    &\mathrm{N2} = (x-8.82)/0.49, \\
    &\mathrm{O3} = 43.9836 - 21.6211x + 3.4277x^2 - 0.1747x^3, \\
    &\mathrm{O3N2} = (8.97 - x)/0.39,
\end{align}
where $x = 12 + \rm \log (O/H)$.
We adopt the \citet{Bian2018} calibrations for consistency with the reference field galaxy measurements \citet{Sanders2020, Li2023}.

We follow \citet{Sanders2021} and derive metallicities by minimizing
\begin{equation}
    \chi^2 = \sum_{i}\frac{(R_\mathrm{obs, i} - R_\mathrm{cal, i})^2}{\sigma_\mathrm{obs, i}^2 + \sigma_\mathrm{cal, i}^2},
\end{equation}
where $R_\mathrm{obs}$ is the observed line ratio, $R_\mathrm{cal}$ is the line ratio predicted by each calibration with a given metallicity, $\sigma_\mathrm{obs}$ is the uncertainty of the line ratio, and $\sigma_\mathrm{cal}$ is the dispersion of each calibration. As \citet{Bian2018} performed fitting on stacked spectra and provided no measured calibration scatters, we simply set $\sigma_{\rm cal}$ as 0.
The \citet{Bian2018} calibrations are extrapolated to $7.8 < \rm 12 + \log (O/H) < 8.8$ ($7.8 < \rm 12 + \log (O/H) < 8.4$ for the sample in \citet{Bian2018}) to cover the high-mass galaxies in the MQN01 sample.
For galaxies with undetected $\rm H\beta$ or [NII], or where [OIII] falls within a spectral coverage gap (see ID: 3172 in Fig.\ref{spec_image_full}), only the available diagnostic is used.
We generate 1000 realizations on emission line fluxes for each galaxy and adopt the median value of the resulting distribution, with uncertainties from 16 and 84 percentile.

\subsection{Sample and comparison with previous observations and field galaxies}
\label{sample_selection}

    \begin{figure}
   \centering
   \includegraphics[width = \columnwidth]{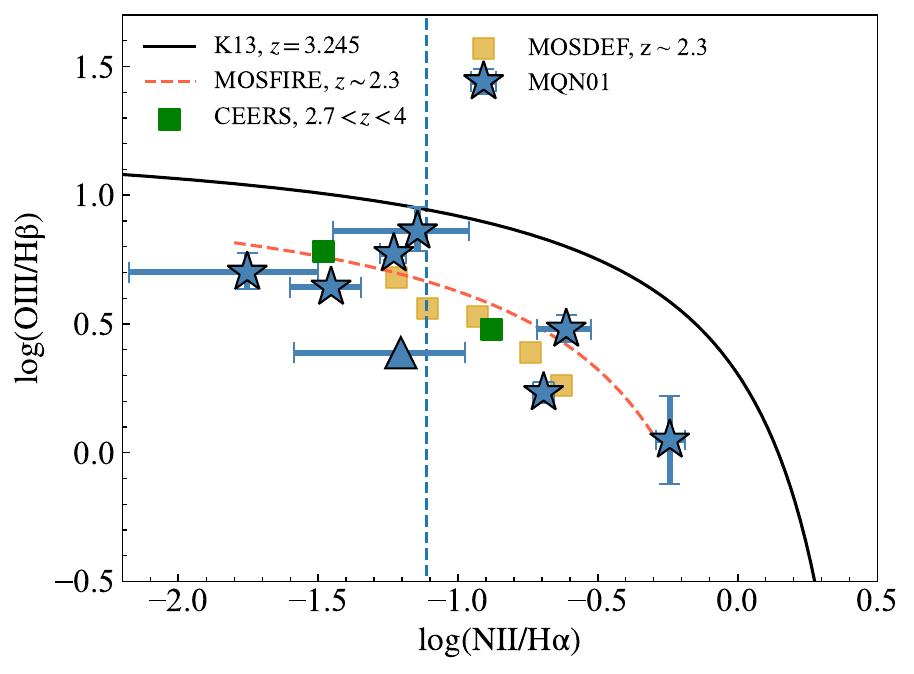}
   \caption{The BPT diagram for the MQN01 galaxy sample. The classification curve is from \citet{Kewley2013} with $z = 3.245$. Galaxies are shown as blue stars. For one galaxy (ID: 6005) with no $\rm H\beta$ detection, a 3$\sigma$ upper limit is adopted, resulting in a lower limit for $\rm [OIII]\lambda5008/H\beta$, shown as a blue triangle.
   The galaxy whose [OIII] falls within the spectra detection gap (ID: 3201, see Fig.\ref{spec_image_full}) is shown as the vertical dashed line at its $\rm [NII]\lambda 6585/H\alpha$ value.
   The distribution of the stacked CEERS $2.7 < z < 4.0$ sample \citep{Shapley2023}, stacked MOSDEF $z \sim 2.2$ \citep{Sanders2021}, and the best fit relation for the KBSS-MOSFIRE sample at $z\sim 2.3$ \citep{Steidel2014}, are also included.}
    \label{BPT}%
    \end{figure}

\begin{figure*} 
    \centering
    \begin{subfigure}{0.48\textwidth}
        \includegraphics[width = \columnwidth]{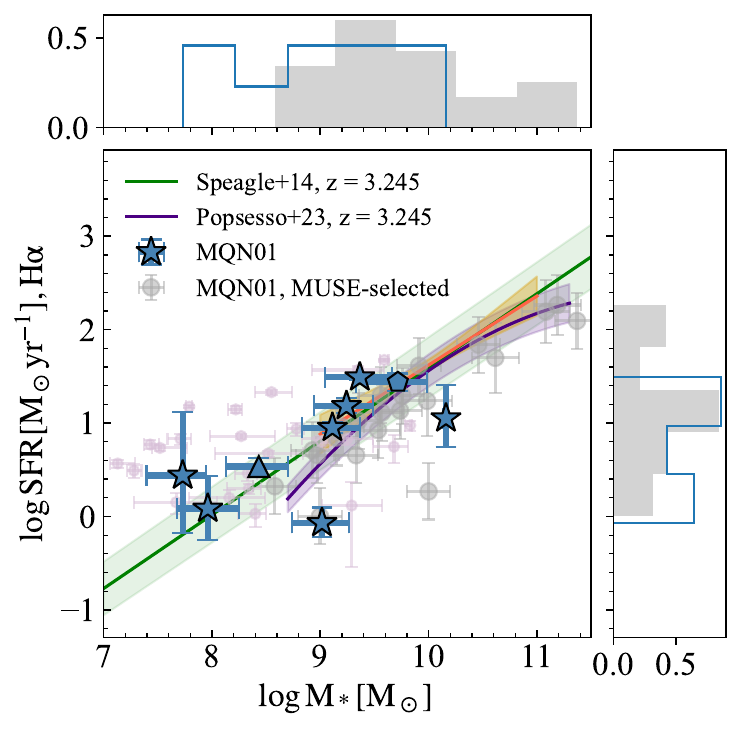}
    \end{subfigure}
    \begin{subfigure}{0.5\textwidth}
        \includegraphics[width = \columnwidth]{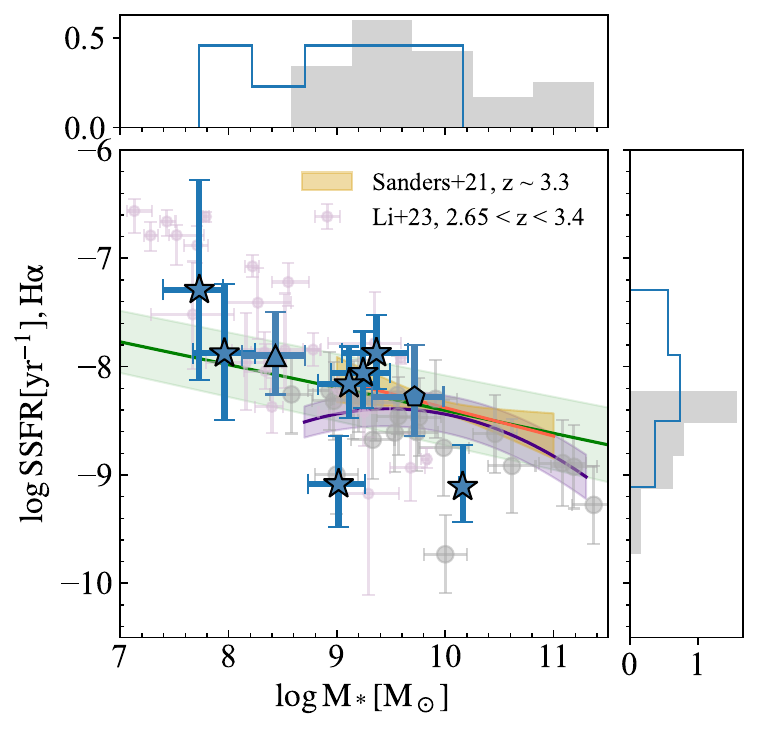}
    \end{subfigure}
    \caption{Star formation rate (SFR) and specific Star formation Rate (SSFR) versus stellar mass for the MQN01 sample. 
    SFR is calculated by $\rm H\alpha$ emission with aperture and dust correction. The galaxy sample is shown as blue stars.
    %The Big Wheel galaxy is shown as a blue diamond. 
    For the galaxy with no $\rm H\beta $ detection (ID: 6005), the SFR without dust correction is adopted as a lower limit, shown as a blue triangle.
    For the galaxy with no JWST photometry (ID: 3201),
    aperture correction calculated from HST F814W is adopted and is shown as a blue pentagon.
    SFR and stellar mass derived from SED fitting for a larger sample of MQN01 galaxies \citep{Galbiati2025} are shown as gray circles.
    The histograms on the top and right in both panels show the distribution of the SFR, SSFR and stellar mass for MQN01 sample in this work (blue) and that from \citet{Galbiati2025} (gray).
    For comparison, the best-fit SFR-$\rm M_*$ relation for the MOSDEF sample with median $z \sim 3.3$ \citep{Sanders2021} is shown as the orange line, with the shaded-region indicating the 1$\sigma$ uncertainty of the fitting parameters. SFR and SSFR versus stellar mass from \citet{Li2023} are shown as purple circles.
    SFR, SSFR-$\rm M_*$ from \citet{Speagle2014} and \citet{Popesso2023} at $z = 3.245$ are shown as the green lines and purple lines, with 1$\sigma$ uncertainties of the fitting parameters shown as the shaded regions.
    Compared with reference SFMS, the MQN01 sample shows a median offset of $\sim $ 0.07 dex above the referred star formation main sequence \citep{Speagle2014}.}
    %5 galaxies reside on the SFMS, while 4 galaxies are about 0.5 - 1 dex off the SFMS.}
    \label{SFMS}
\end{figure*}

We have checked emission line features for all slit spectra and identified 15 galaxies with clear emission detections within $\sim 1500\rm km/s$ relative to the central QSO. 8 of the 15 galaxies have been identified in the MUSE-selected sample \citep{Galbiati2025}, and 5 of 15 have X-ray emission \citep{Travascio2025}.
In the 10 galaxies without X-ray detection, one has been identified as quiescent (Wang et al, in prep.).
% A summary of... samples can be
Due to the known complications in the gas excitation mechanisms with AGNs in presence and the resulting uncertainties in metallicity measurements, AGN hosts are excluded in the sample for analysis.
% 9 + 2 AGN + AGN triplets(2 or 3?) + Big Wheel + Red potato
We finally identified a sample of 9 star-forming galaxies (SFGs).
The distribution of the parent (15 galaxies) and identified (9 galaxies) sample is shown as colored circles in Fig. \ref{sky_region}.
The BPT diagram of the final sample is shown in Fig. \ref{BPT}.

In the final sample of SFGs, 3 of 9 are included in the MUSE-selected sample, and 1 of the remaining 6 is included in the ALMA sample.
%Compared with the MUSE-selected sample, the new JWST-selected sample extends more toward low-mass galaxies.
The distribution of the MQN01 JWST sample on the star formation main sequence is presented in Fig. \ref{SFMS}, with the MUSE-selected sample plotted for comparison.
Relative to the MUSE sample, the MQN01 sample extends toward lower stellar masses and systematically lower SFRs, while the specific SFRs (sSFRs) remain comparable. 
This is due to the high fraction of AGNs at the high-mass end in the MQN01 field \citet{Travascio2025}, whereas AGNs are excluded in this work.
For star forming galaxies with $M < 10^{10.5} M_\odot$, the two samples are broadly consistent.
% ALMA:
% 9
% BW, 1148, none, 3025, none, none, none, none, 3001
% L03, 4, 5, 6, 7, 8, 9

To place the MQN01 sample into context, we also compare the MQN01 sample with several widely used SFMS relations and other samples at similar redshifts, as shown in Fig. \ref{SFMS}.
We adopt the parameterizations of SFMS from \citet{Speagle2014} and \citet{Popesso2023} as general benchmarks, and include the field-galaxy samples at similar redshifts from \citet{Sanders2021} and \citet{Li2023} for direct observational comparison.
% We also present the SFMS from Speagle14 and Popesso23 for reference, and SFMS for the field galaxies at similar redshifts from \citet{Sanders2021} and \citet{Li2023} for comparison.
The MQN01 SFGs have comparable SFR with the reference SFMSs.
% We use a Monte Carlo method to estimate the SFR difference with reference star-forming main-sequence (SFMS) relations of the MQN01 sample at a fixed stellar mass. For each galaxy, we generate 10,000 realizations of stellar mass and SFR from the observed values and measurement errors. For the reference relations, we generate 10,000 realizations of the relation parameters and calculate the corresponding SFR values.
% The resulting distributions are used to quantify the SFR offsets between the MQN01 sample and the reference relations. On average, the MQN01 sample lies $0.07\pm 0.163$ above the SFMS in \citet{Speagle2014}. At $M > 10^{9} M_\odot$, the SFR offsets are $-0.085\pm 0.171$\footnote{Positive values mean above the relative SFMS, while negative values mean low the relative SFMS.} relative to \citet{Speagle2014}, and $0.07\pm 0.153$ relative to \citet{Popesso2023}, and $-0.14\pm 0.129$ relative to \citet{Sanders2021}. The MQN01 SFGs have comparable SFR with the reference SFMSs with the differences.. smaller than 1sigma systematic difference...

\section{Line ratios and metallicity versus stellar mass of MQN01}
\label{result}
In this section, we present our major results of galaxy chemical properties, including emission line ratios and metallicity of the MQN01 star forming galaxy sample.

\subsection{Line ratios}
\begin{figure*} 
    \centering
    \begin{subfigure}{0.49\textwidth}
        \includegraphics[width = \columnwidth]{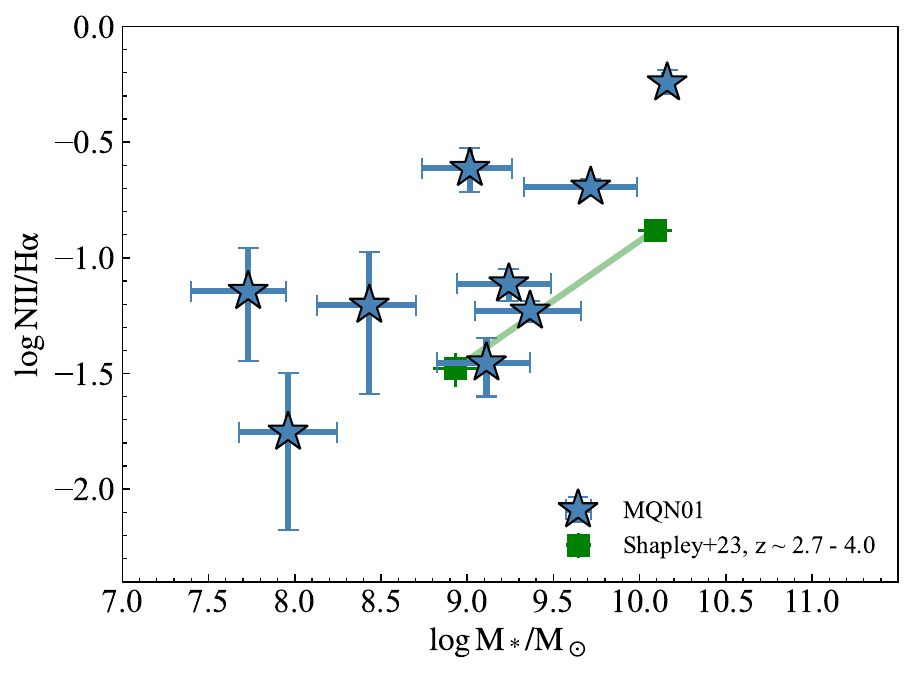}
    \end{subfigure}
    \begin{subfigure}{0.49\textwidth}
        \includegraphics[width = \columnwidth]{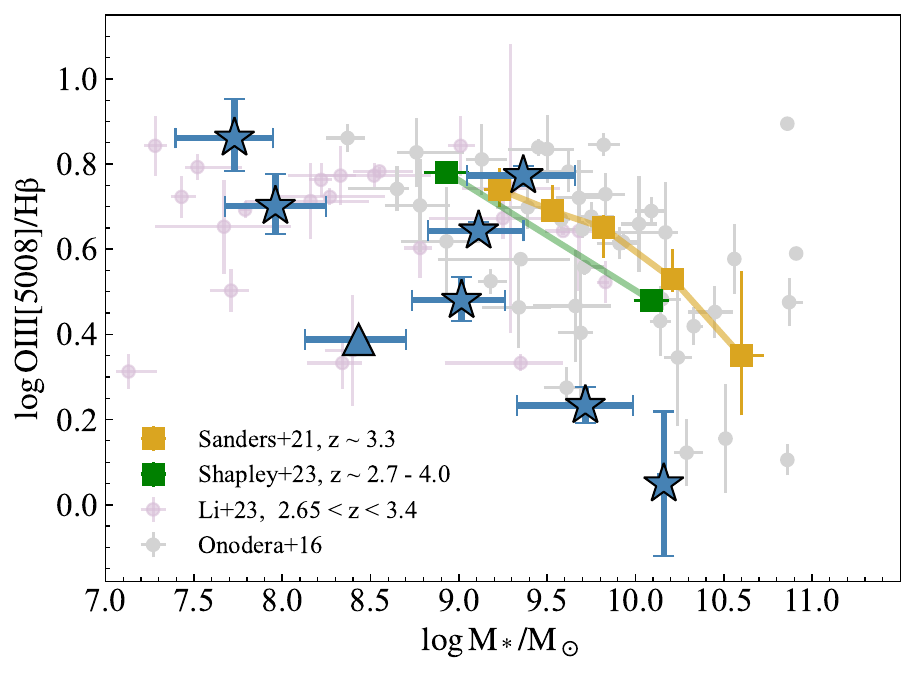}
    \end{subfigure}
    \caption{Emission line flux ratio versus stellar mass for the MQN01 sample. The sample is marked as blue stars.
    %and the Big Wheel is shown as blue diamond.
    Left: $\log $ [NII]$\lambda$6585/$\rm H\alpha$ versus stellar mass. Reference data from stacked spectra of the CEERS sample with $2.7 < z < 4.0$ \citep{Shapley2023} are shown as green squares. Right: $\log$ [OIII]$\lambda5008$/$\rm H\beta$ versus stellar mass. The galaxy without $\rm H\beta$ detection is shown as a triangle at the lower limit of [OIII]$\lambda5008$/$\rm H\beta$ based on the 3$\sigma$ detection. The datasets from \citet{Onodera2016, Sanders2021, Li2023, Shapley2023} are included for comparison. The individual data points are shown in filled circles, with stacked values shown as squares.
    Compared with the reference datasets, the MQN01 galaxies show on average higher [NII]$\lambda6585$/$\rm H\alpha$ and lower [OIII]$\lambda5008$/$\rm H\beta$.}
    %For $\rm \log M_*/M_\odot > 9$, three galaxies reside on the median trend of \citet{Sanders2021} and \citet{Shapley2023}, while three galaxies are more above/below the sequence.}
    %The Big Wheel is more consistent with the median sequence in both panels.}
    \label{fluxratio_M}%
\end{figure*}

We begin by investigating the relations between emission line ratios and stellar masses.
Fig. \ref{fluxratio_M} shows [NII]$\lambda6585$/$\rm H\alpha$ and [OIII]$\lambda5008$/$\rm H\beta$ versus stellar mass for the MQN01 sample.
The galaxy with no $\rm H\beta$ detection (ID: 6005) is plotted as a triangle with 3$\sigma$ upper limit of $\rm H\beta$ in the right panel of Fig. \ref{fluxratio_M}.
For comparison, we include the relations for field galaxies at similar redshifts. We include the stacked CEERS sample \citep{Shapley2023} for [NII]$\lambda6585$/$\rm H\alpha$], as observations at $z > 3$ are rare due to limited wavelength coverage, and we include the samples from \citet{Onodera2016}, \citet{Sanders2021} and \citet{Li2023} for [OIII]$\lambda5008$/$\rm H\beta$, which is more available at $z > 3$.

In the MQN01 sample, [NII]$\lambda6585$/$\rm H\alpha$ increases while [OIII]$\lambda5008$/$\rm H\beta$ decreases with stellar mass.
[NII]$\lambda6585$/$\rm H\alpha$ and [OIII]$\lambda5008$/$\rm H\beta$ serve as metallicity indicators.
[NII]$\lambda6585$/$\rm H\alpha$ increases with metallicities, while [OIII]$\lambda5008$/$\rm H\beta$ first increases and then decreases with metallicities when $\rm 12 + \log (O/H) \gtrsim 8$ \citep{Maiolio2008, Curti2017, Bian2018, Sanders2020}. This metallicity regime is a reasonable estimate for our sample given its stellar mass range.
Therefore, the observed trends of the two flux ratios are consistent with a metallicity increase toward higher stellar masses, consistent with the well established mass-metallicity relation.

% consistent with a metallicity increase toward higher masses.
% The MQN01 galaxies show consistent line ratio trends with higher metallicity with increasing stellar mass, with [NII]$\lambda6585$/$\rm H\alpha$ increasing and [OIII]$\lambda5008$/$\rm H\beta$ decreasing. 
Compared with field galaxies at similar redshifts, MQN01 galaxies cover a similar range in each emission-line ratio but show a systematic offset towards higher metallicities (higher [NII]$\lambda6585$/$\rm H\alpha$ and lower [OIII]$\lambda5008$/$\rm H\beta$).
This is clearer at $\rm \log M_*/M_\odot > 9$, where MQN01 galaxies consistently tend to lie above the [NII]/H$\alpha$ and below the [OIII]/H$\beta$ values of stacked spectra from \citet{Shapley2023} and \citet{Sanders2021} at fixed mass.
At $\rm \log M_*/M_\odot < 9$, there is no direct comparison with field galaxies for [NII]$\lambda6585$/$\rm H\alpha$ and limited datasets for [OIII]$\lambda5008$/$\rm H\beta$.
MQN01 galaxies tend to lie above the extrapolated [NII]/H$\alpha$–mass relation from \citet{Shapley2023} at $\rm \log M_*/M_\odot < 9$, while their [OIII]/H$\beta$ ratios are broadly consistent with the measurements of \citet{Li2023}. Given the limited statistics at the low-mass end, it remains difficult to draw firm conclusions for the full mass range.
Nevertheless, the overall trend suggests that MQN01 star-forming galaxies tend to have relatively higher metallicities than their field counterparts, especially at high-mass end.

\subsection{Mass-metallicity relation}

\begin{figure*} 
    \centering
    \includegraphics[width = 1.5\columnwidth]{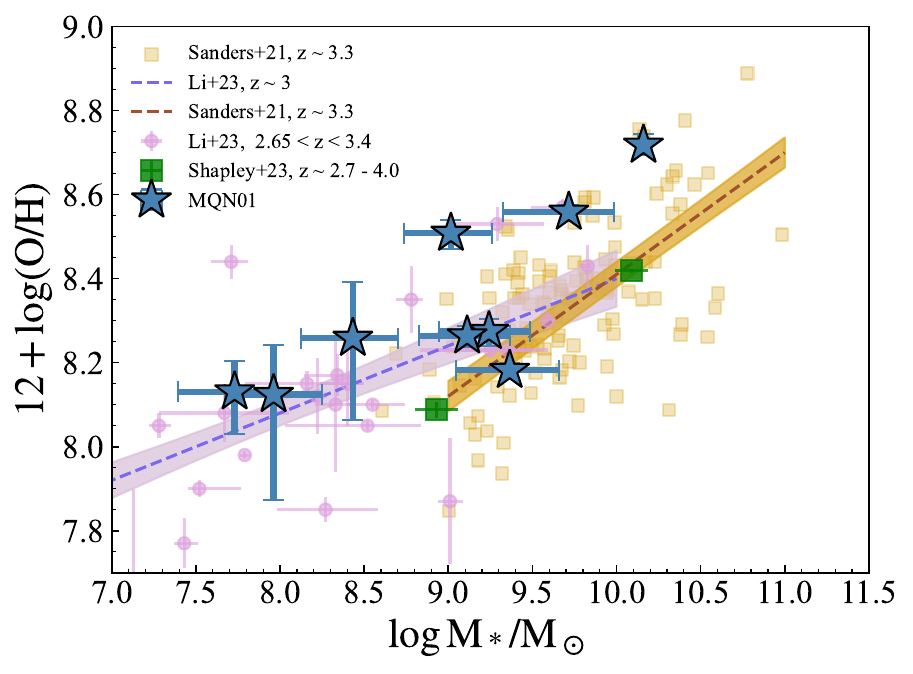}
    \caption{Mass-metallicity relation for the MQN01 sample. The MQN01 galaxies are shown as blue stars.
   The reference datasets at similar redshifts from \citet{Sanders2021, Li2023, Shapley2023} are included.
   The reference datasets are from \citet{Sanders2021, Li2023, Shapley2023}, where the individual galaxies are shown as filled circles, with the best-fit relations shown in lines and shaded regions representing 1$\sigma$ uncertainties.
   The individual data points of the sample in \citet{Sanders2021} are extracted using WebPlotDigitizer.
   The CEERS sample \citep{Shapley2023} is shown in green circles, with metallicities calulated using N2, O3 and O3N2 with \citet{Bian2018} calibrations.
   Even though the median best fit relations for different reference datasets differ, the distribution of individual points overlap.
   Consistent with Fig. \ref{fluxratio_M}, the MQN01 galaxies show on average higher metallicity. For $\rm \log M_*/M_\odot > 9$ where metallicities have less uncertainties, three MQN01 galaxies reside close to the upper envelope of the reference datasets, while three galaxies are more consistent with the median trend from \citet{Sanders2021}.}
    \label{MZR}
\end{figure*}

 \begin{figure}
   \centering
   \includegraphics[width = \columnwidth]{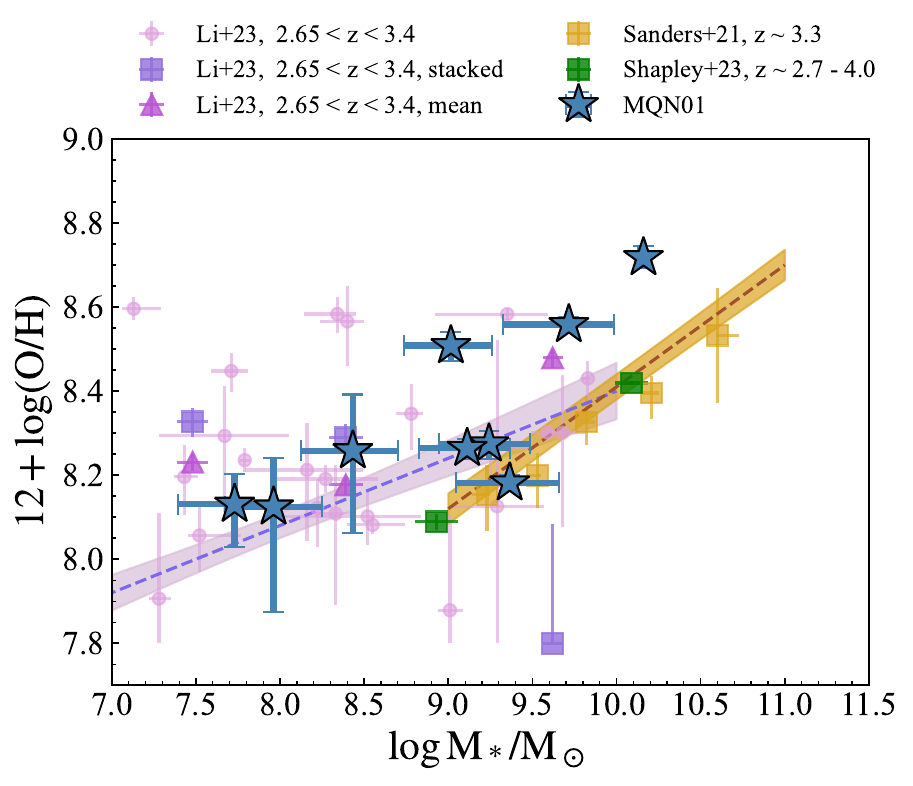}
   \caption{Same as Fig. \ref{MZR}, but the metallicity for reference datasets are calculated in the same way with the MQN01 sample. For the CEERS sample \citep[Green squares, ][]{Shapley2023}, the metallicities are calculated using N2, O3 and O3N2 with \citet{Bian2018} calibrations, while for the samples in \citet{Li2023} (purple circles and squares) and \citet{Sanders2021} (brown squares) the metallicities are calculated using O3 only.
   }
    \label{MZR_o3}%
    \end{figure}

\begin{figure*} 
    \centering
    \begin{subfigure}{\columnwidth}
        \includegraphics[width = \columnwidth]{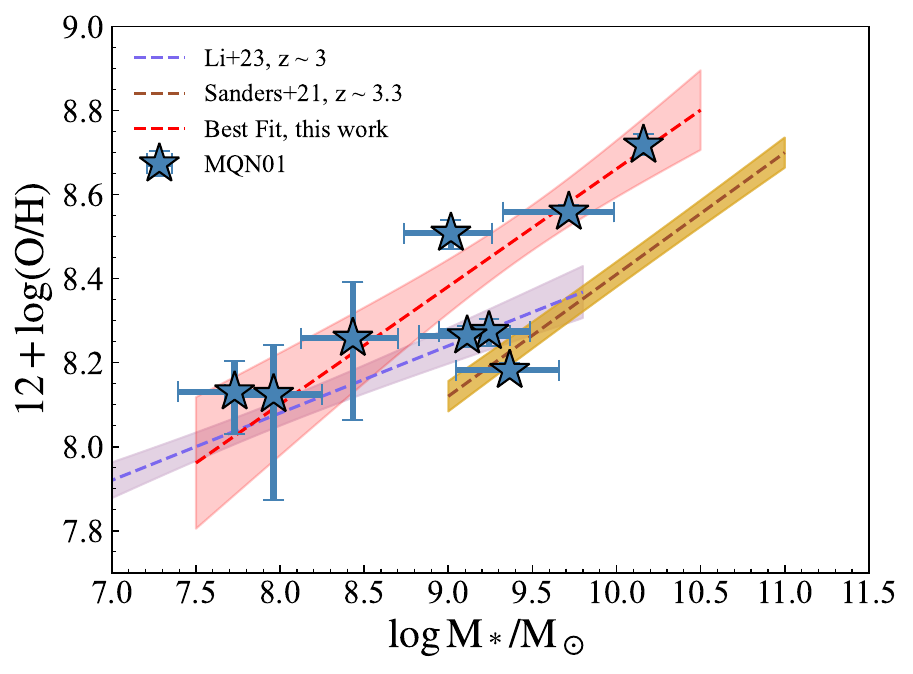}
    \end{subfigure}
    \begin{subfigure}{\columnwidth}
        \includegraphics[width = \columnwidth]{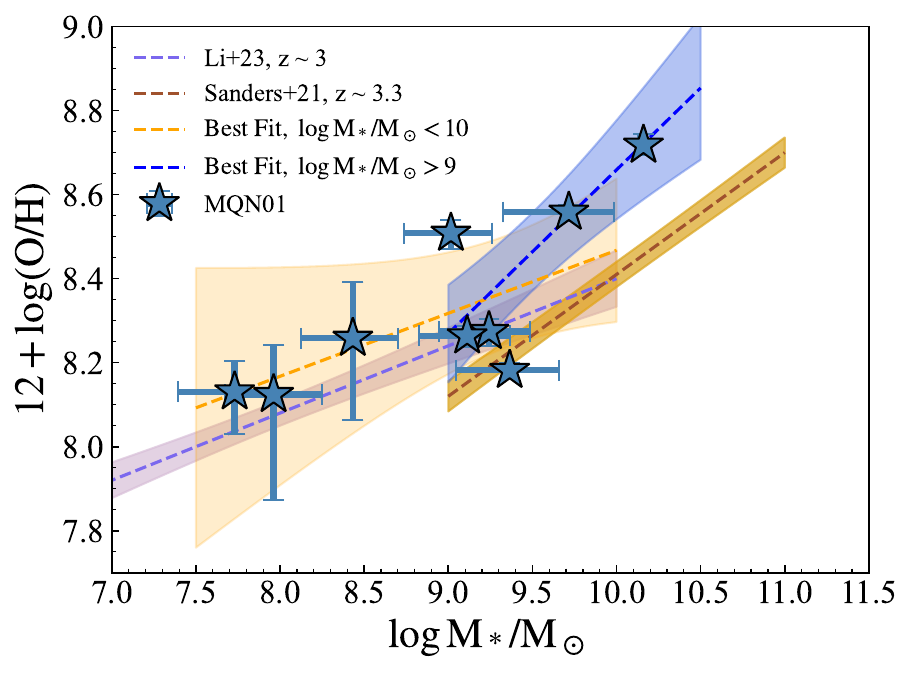}
    \end{subfigure}
    \caption{Best-fit mass-metallicity relation for the MQN01 sample. The MQN01 galaxies are shown as blue stars.
   The reference relations at similar redshifts from \citet{Sanders2021, Li2023} are included.
   Left: The best-fit relation with 1$\sigma$ uncertainties of the total MQN01 sample.
   Right: The best-fit relations with 1$\sigma$ uncertainties of the MQN01 sample with $\rm \log M_*/M_\odot < 10$ and $\rm \log M_*/M_\odot > 9$, to keep consistent with the mass range of \citet{Li2023} and \citet{Sanders2020}.
   }
   
    \label{MZR_lines}
\end{figure*}

We present the mass-metallicity relation (MZR) for the MQN01 sample in Fig. \ref{MZR}, offering one of the first views of the MZR for protocluster galaxies at $z \sim 3$. For comparison, we include reference results for field galaxies from \citet{Sanders2021} and \citet{Li2023}, which are shown as both median trends and individual data points. We also include the metallicity for binned spectra of CEERS at $2.7 < z < 4.0$, with N2, O3 and O3N2 from \citet{Shapley2023}, calculated in the same way with the MQN01 sample.

The MQN01 sample shows a clear increasing metallicity trend versus stellar mass.
% We notice that the median trends of reference data are not consistent. \citet{Li2023} is a bit higher than \citet{Sanders2021}. Even so, the distributions of scatter points for the three datasets are largely overlapped.
Compared with the reference data, the MQN01 galaxies show averagely higher metallicities at fixed stellar mass, consistent with what is seen in flux ratios in Fig. \ref{fluxratio_M}.
At $\rm \log M_*/M_\odot > 9$, the MQN01 galaxies show a mean offset of $0.18 \pm 0.036$ dex above to the median relation presented from \citet{Sanders2021}.
At $\rm \log M_*/M_\odot < 9$, the distribution of MQN01 metallicities is closer to that of the \citet{Li2023} sample, with a mean offset of only $0.03 \pm 0.09$ dex. The latter difference is modest and statistically insignificant within $1\sigma$ uncertainties.

The metallicity measurements are sensitive to the calibration method. We adopt the \citet{Bian2018} calibration for consistency with the reference data.
However, the specific set of emission line diagnostics may vary between samples due to differences in wavelength coverage and data quality.
To investigate the potential impact of these differences, we calculate metallicities for the \citet{Sanders2021} and \citet{Li2023} samples with O3, which is the only available line ratio overlapping with what used in this work.
The re-calculated mass-metallicity relations are shown in Fig. \ref{MZR_o3}.
For the \citet{Sanders2021} sample, the O3-based metallicities are highly consistent with the publishsed values with slight offset.
For the \citet{Li2023} sample, however, the recalculated metallicities reach higher values and the measurements from stacked spectra show even a decreasing trend with stellar mass, which is inconsistent with the well-established MZR.
This may reflect methodological details in the stacking procedure and is beyond the scope of this work.
As an additional cross-check, we derived binned O3 metallicities by summing the reported line fluxes. These values, shown as purple triangles in Fig. \ref{MZR_o3}, 
are more consistent with the established MZR but lie systematically above the published relation of \citet{Li2023}, and they show no clear difference from the MQN01 distribution. 
We also notice that the O3-based metallicities for the \citet{Li2023} sample exhibit larger scatter compared to the published values. 
The inconsistencies between the O3-based and published metallicities possibly indicate the limited constraining power of O3 on metallicities or uncertainties in measuring weak $\rm H\beta$ emission, while a detailed assessment is beyond the scope of this work.
Overall, these O3-based recalculations provide a sanity check, while in the following we adopt the published relations in Fig.~\ref{MZR} as the common reference for statistical comparison, in order to remain directly comparable with earlier work.

For statistical comparison with field galaxies, we model the mass-metallicity relation of the MQN01 sample with a linear form, 
\begin{equation}
    \mathrm {\log (O/H) + 12} = k(\log M_* - \log M_0) + Z_0,
\end{equation}
where $\log M_0$ is the mean mass weighted by the inverse variance of the metallicities.
The best fit parameters and $1\sigma$ uncertainties are derived from MCMC fitting.
As the field galaxy samples cover different mass ranges, we also fit the galaxy sample subsets with the same mass coverage as the field galaxy samples.
The best fit relations with uncertainties for the MQN01 sample and for reference relations are shown in Fig. \ref{MZR_lines}, and the best fit parameters are summarized in Table \ref{para_table}.
% For a better comparison with the field galaxy samples, which cover different mass ranges, we also fit the MQN01 sample for mass < 9 and mass > 9 separately. 
Compared with \citet{Sanders2021}, the MQN01 MZR has a consistent slope ($\Delta k = -0.01\pm0.08$) and a systematic offset of $0.25\pm0.073$ toward higher metallicity.
Restricting to $\log(M_\ast/M_\odot)>9$, the MQN01 slope is slightly steeper with similar offsets.
% ; the metallicity offset is smaller at the low-mass end ($0.205\pm0.086$ dex at $\rm \log(M_\ast/M_\odot)=9.5$) and increases to $0.295\pm0.086$ dex at $\rm \log(M_\ast/M_\odot)=10.5$.
When compared with \citet{Li2023}, restricting to $7.5 <\log(M_\ast/M_\odot)<10$, the best-fit relation shows larger scatter. The average offset remains toward higher metallicity ($0.09 \pm 0.26$) with consistent slopes ($\Delta k = -0.01\pm0.17$), while the significance is below $1\sigma$. The average difference would become even less significant if O3-based metallicities are used (see Fig. \ref{MZR_o3}).
The statistical analysis suggests a systematic metallicity excess in the MQN01 galaxies relative to the field galaxy samples, most pronounced at the high-mass end. At lower masses, a similar tendency is present while with limited significance.

\begin{table*}
    \centering
    \begin{tabular}{lllllll}
        \hline
        \hline
        Sample & Range of $ \log M_*/M_{\odot}$ & $k$ & $\log M_0$ & $Z_0$ & $\log M_{\rm ref}$   &  $Z$ at $\log M_{\rm ref}$\\
        \hline
        \multirow{3}{*}{MQN01}
         & (7.5, 10.5) & $0.28\pm 0.08$ & 9.46 & $8.51\pm 0.05$ & 10 & $8.66\pm 0.07$ \\
         & (9.0, 10.5) & $0.39\pm 0.14$ & 9.49 & $8.46\pm 0.09$ & 10 & $8.66\pm 0.11$ \\
         & (7.5, 10.0) & $0.15\pm 0.17$ & 9.35 & $8.37\pm 0.13$ & 8  & $8.17\pm 0.26$ \\
         \hline
         \citet{Sanders2021} & (9.0, 11.0) & $0.29\pm 0.02$ & 10 & $8.41\pm 0.03$ & 10 & $8.41\pm 0.03$\\
         \hline
         \citet{Li2023} & (6.5, 10.0) & $0.16\pm 0.03$ & 8 & $8.08\pm 0.03$ & 8 & $8.08\pm 0.03$ \\
        \hline
    \end{tabular}
    \caption{Best-fit mass-metallicity relation parameters for the MQN01 sample and the reference samples. The pivot mass ($\log M_0$) for the MQN01 sample is the mean mass weighted by the inverse variance of the metallicities, where asymmetric errors were symmetrized by averaging the upper and lower values.
    The metallicities of the MQN01 sample at the pivot masses of the reference samples are also listed for comparison.
    We set $10^{10} M_\odot$ and $10^{8} M_\odot$ as the reference stellar masses ($M_{\rm ref}$) for the high and low mass regimes respectively, matching the $M_0$ in the parameterized MZR fits of the field galaxy samples, for a direct comparison.
    }
    \label{para_table}
\end{table*}

\section{Discussion}
\label{dis}
\subsection{Environmental effect in metal enrichment}

In the previous sections, we have shown the mass-metallicity relation for the MQN01 sample.
The MQN01 galaxies show a tendency towards higher metallicities than field galaxies, especially at higher masses, with less significance at low mass. These differences may indicate environmental effects on metal enrichment with mass dependence.

Previous observations have suggested that galaxies in the overdense environments may assemble their stellar mass more rapidly or at earlier times than field galaxies \citep{Galbiati2025}.
The earlier star formation in protoclusters therefore leads to earlier metal enrichment, while field galaxies experience more delayed metal enrichment. This is consistent with the high fraction of massive galaxies and AGN in MQN01, which indicates accelerated growth. 
%and feedback.
%At lower masses,  metallicities are more comparable to those of the field, possibly because ongoing accretion of pristine gas dilutes enrichment. 
%galaxies in protoclusters are at earlier stage of star formation with gas accretion, which explains the comparable metallicities with field galaxies.
Other mechanisms, like gas stripping and galaxy interactions that lead to metal-poor gas loss and metal enhancement, may also play a role. However, they would be expected to enhance metallicities especially in low-mass systems, which is not observed here.
We also notice that the slope of the MZR of MQN01 given by fitting within different mass regions are consistent with the reference field galaxies, possibly suggesting that the environmental effects primarily shift the normalization rather than the shape of the relations at $z \sim 3$. At low masses, the slope of MZR is shallower than that of the high mass end, possibly due to different feedback mechanisms at different mass ranges \citep{D2011, Li2023}.
However, when fitting the full sample, the slope is steeper and more consistent with the MZR for massive galaxy samples. Given that the current sample is still quite limited, further observations are required to better constrain the low-mass regime and to establish whether environment leaves a robust imprint on the MZR at $z \sim 3$.

Observations of protoclusters at $ z > 3$ are still sparse. As summarized in Sec. \ref{intro}, at $z \sim 2$ observations have reported inconsistent results. Some have reported metallicity deficit on the high mass end and enhancement at low mass end, which are interpreted as a combined effect of metal enrichment through feedback and dilution by cold gas accretion \citep{Kulas2013, Kacprzak2015, Chartab2020, Sanders2021, Wang2022}.
Some other observations have reported metallicity enhancement which are interpreted as suppressed cold gas supply by shock heated protocluster haloes and more efficient metal-enriched gas recycling \citep{Shimakawa2014, Shimakawa2015,P2023}.
At higher redshifts, $5 < z < 7$, \citet{Li2025} has reported metallicity enhancement in protoclusters.

We stress, however, that different systems are identified as ``protoclusters'' based on different criteria and, in some cases, even modest overdensities on large scales are considered ``protoclusters'' in the literature.
The observations and analysis of MQN01 add to the limited observational constraints on the MZR of protocluster galaxies at $z \sim 3$, for a system that shows the highest overdensity in terms of continuum-selected star forming galaxies, CO emitters and AGN, at least in the central area of 4 arcmin$^2$ covered by JWST observations and ALMA \citep{Galbiati2025, Pensabene2024, Travascio2025}. Unfortunately, different galaxy selections in different fields hamper the possibility of a detailed comparison based on overdensities.
%
%providing a data point between the regime of $z \sim 2$ and $z > 5$. 
The higher metallicities seen in MQN01 possibly suggest that metallicity enhancement may still be present above a given overdensity of galaxies at $z \sim 3$.
%As MQN01 is observed to be one of the most overdense protoclusters, 
Further observations of protoclusters across a wider range of overdensities and redshifts would be essential to further understand the environmental impact on MZR and its redshift evolution.

\subsection{The fundamental metallicity relation}
    \begin{figure}
   \centering
   \includegraphics[width = \columnwidth]{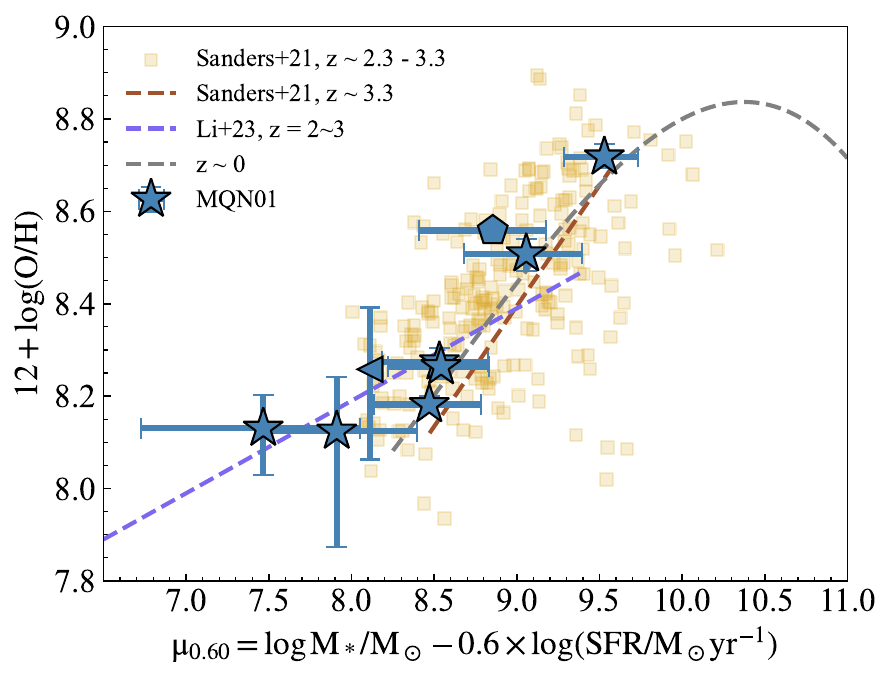}
   \caption{Gas-phase metallicity versus $\mu_{\alpha}$, known as the fundamental metallicity relation. The x-axis $\mu_{\alpha} = \log M_*/M_\odot - \alpha \log (\rm SFR[M_\odot yr^{-1}])$. We adopt the value of $\alpha = 0.60$ from \citet{Sanders2021}. The MOSDEF sample individual galaxies at $z \sim 2.3$ and $z \sim 3.3$ are shown in brown squares (extracted using WebPlotDigitizer). The FZR calculated by the best-fit MZR and best-fit SFMS from \citet{Sanders2021} is shown as the brown dased line.
   The gray dashed line shows the best-fit cubic function to the stacks of $z \sim 0$ datasets, as presented in Equation 10 in \citet{Sanders2021}. The MOSDEF sample with $z \sim 3.3$ shows good agreement with the FZR at $z \sim 0$.
   The purple dashed line shows the best-fit relation of \citet{Li2023}.
   The MQN01 sample is shown as blue stars. For the galaxy with no $\rm H\beta$ detection (ID: 6005), the SFR without dust correction is adopted and shown as the blue triangle at the upper limit of $\mu_{0.60}$. The galaxy with no JWST photometry is marked as pentagon.
   }
    \label{FMR}%
    \end{figure}

In this subsection we further investigate the fundamental metallicity relation (FZR), which combines stellar mass and star formation. FZR is defined as a function of $\mu_{\alpha}$ instead of mass, where $\mu_{\alpha} = \log M_*/M_\odot - \alpha \log (\rm SFR[M_\odot yr^{-1}])$. 
We adopt the value of $\alpha = 0.60$ from \citet{Sanders2021}. The resulting FZR is shown in Fig. \ref{FMR}. Same as in Figs. \ref{SFMS} and \ref{fluxratio_M}, the galaxy with no $\rm H\beta$ detection (ID: 6005) is marked as a triangle with an upper limit of $\mu_{0.60}$, and the galaxy with no JWST photometry (ID: 3201) is marked as a pentagon. The MOSDEF sample is included for comparison and is shown as brown squares. The FZR fitted with a $z \sim 0$ sample presented in \citet{Sanders2021} is shown as the dashed curve.
\citet{Sanders2021} has pointed that the FZR has weak dependence on redshift. The distribution of MOSDEF samples with $z \sim 2.3$ and $z \sim 3.3$ show both good agreement with the $z \sim 0$ curve. The best fit relation of the \citet{Li2023} sample is also included.

The MQN01 galaxies on the FZR tend to lie to the left of the reference relations from \citet{Sanders2020} with higher metallicities, consistent with what is seen on MZR.
However, the difference is less significant and within $1\sigma$ measurement uncertainties.
We also notice that galaxies either on or above the reference MZR of field galaxies \citep{Sanders2020} are broadly consistent with the field galaxies on FZR.
Overall, the MQN01 sample follows similar distributions and slopes on the FZR with the field galaxy samples, both at low and high $\mu_\alpha$.
The better consistency between protocluster galaxies and field galaxies on FZR than MZR suggests that the environmental effect may not sufficiently affect the fundamental regulation between star formation and metal enrichment, which is generally attributed to complex balance between gas inflows, outflows and feedback \citep{Lilly2013}.
These interpretations still need further investigation, as FZR is sensitive to systematic uncertainties, including the choice of $\alpha$, calculation of SFR and metallicity calibrations (see also discussions in \citet{KC2025}).
% The composite uncertainties may contribute to the scatters observed on FZR, making it hard to .. the enviornmental effects contribute to ... sample.. 
We therefore keep the interpretation open until further observations with consistent calibrations are available.

\subsection{Sample bias and model uncertainties}

Even though we see statistical metal enhancement in MQN01 compared with field galaxies, the limited sample size (9 galaxies) prevents us from drawing firm conclusions with high significance.
The MZR for MQN01 is derived from fitting individual scattered points, while the reference relations are based on stacked spectra.
We notice that when comparing with distributions of scatter points instead of stacked values, the MQN01 galaxies do not show a clear excess in metallicities relative to the most metal-rich field galaxies.
Cautions need to be taken whether the relations are biased by a few galaxies rather than reflecting the real systematic differences.

Furthermore, differences in sample selection criteria may introduce additional biases. For example, the sample from \citet{Li2023} is selected based on strong OIII emission, which may tend to favor galaxies with particular ionization conditions given the non-monotonic dependence of O3 on metallicities. Galaxies with extreme low or high metallicities may be missed.
In addition, high-metallicity galaxies may be more dust-obscured with weaker emission lines.
Methodological differences in fitting and stacking may also introduce systematic uncertainties.

% We notice that the sample presented in this work is limited to star-forming galaxies with emission-line detections. Massive galaxies in MQN01 are AGNs.
% While spectra of some AGN hosts in MQN01 are also available, their metallicities are challenging to constrain due to ionization effects. A detailed analysis of ionization states of AGNs in MQN01 will be helpful to further understand the metal enrichment in overdense environments.

The MZR and FZR are sensitive to measurement uncertainties.
The stellar mass measurements are from SED fitting. Even though we have adopted similar assumptions in the fitting procedure, differences in photometric coverage can affect the constraints. 
Furthermore, the star formation rates are estimated with different tracers. As $\rm H\alpha$ observations are not always available at $z > 3$, the SFR estimates from \citet{Sanders2020} and \citet{Li2023} are from $\rm H\beta$ emissions and $E(B-V)$ from SED fitting, which add uncertainties and possible systematic differences.
Finally, different choices of metallicity calibrations and adopted diagnostics add further uncertainties. 
As shown in Fig. \ref{MZR_o3}, even though we have adopted the same calibration methods \citep{Bian2018}, the MZR at low mass may be highly dependent on which strong-line diagnostics are used.
Larger samples with homogeneous photometry, consistent SFR tracers, and uniform metallicity calibrations will be essential to establish a more complete description of environmental effects on the MZR and FZR at $z \sim 3$.

% While spectra of some AGN hosts in MQN01 are also available, their metallicities are challenging to constrain due to ionization effects. A detailed analysis of ionization states of AGNs in MQN01 will be helpful to further understand the metal enrichment in overdense environments.
% Overall, more observations are urgently needed to further understand the chemical evolution in different environments and their cosmic evolution.

\section{Conclusions}
\label{conc}
We presented the mass-metallicity relation for a sample of 9 star-forming galaxies in the MUSE Quasar Nebula 01 (MQN01) field, a massive cosmic web node at $z \sim 3.245$, containing one of the largest overdensities of galaxies and AGNs known at these redshifts.
Compared with field galaxies at similar redshifts, the MQN01 sample shows on average $\sim 0.25$ enhanced metallicities on the mass-metallicity relation with similar slopes.
The enhancement is more prominent at higher stellar mass ($\log M_*/M_\odot > 9$) and less significant at lower stellar mass.
The differences between MQN01 and field galaxies are less significant on the fundamental metallicity relation, where metallicity is a function of $\mu_{0.6} = \log M_* - 0.6\log \rm SFR$.
The enhanced metallicities of MQN01 suggest that galaxies in dense environments may experience more efficient or earlier stellar mass growth and metal enrichment than their field counterparts.
These results provide one of the first views of mass-metallicity relations in protoclusters at $z \sim 3$, especially for one of the most overdense environments at $z \gtrsim 3$.
% Firm conclusions are difficult to draw given the limited sample size and uncertainties in metallicity calibrations.
Future observations across larger samples and broader redshift ranges will be important to test whether the features seen in MQN01 are representative of protocluster environments in the early universe, and to further understand cosmic evolution of protoclusters and relevant physical processes.

\begin{acknowledgements}
This work is supported by the European Research Council (ERC) Consolidator Grant
864361 (CosmicWeb) and by Fondazione Cariplo grant no. 2020-0902. XW acknowledges further support by Westlake University and the National Science Foundation of China (Grant No. 11821303 to SM).
This work is based on observations made with the NASA/ESA/CSA James Webb Space Telescope. The data were obtained from the Mikulski Archive for Space Telescopes at the Space Telescope Science Institute, which is operated by the Association of Universities for Research in
Astronomy, Inc., under NASA contract NAS 5-03127 for JWST. These observations are associated with program \#1835. 
Support for program \#1835 was provided by NASA through a
grant from the Space Telescope Science Institute, which is operated by the Association of Universities for Research in Astronomy, Inc., under NASA contract NAS 5-03127.
This research is based on observations made with the NASA/ESA Hubble Space Telescope obtained from
the Space Telescope Science Institute, which is operated by the Association of Universities for 15 Research in Astronomy, Inc., under NASA contract NAS 5–26555. These observations are associated with program 17065. 
This work is also based on observations collected at the European Southern Observatory under ESO programme (110.23ZX).

\end{acknowledgements}

\bibliographystyle{aa} % style aa.bst
\bibliography{ref} % your references Yourfile.bib
\begin{appendix}

% WW: add RA and DEC; keep sig figs of redshift values the same
\section{Measurements of MQN01 sample}
In this section we present a summary Table \ref{prop_table} of the measured and derived properties of galaxies in the sample, including spectroscopic redshifts, stellar mass, star formation rate derived from $\rm H\alpha$ flux and corrected for dust attenuation and slit loss, line ratios ($\rm \log [OIII]\lambda5008/H\beta$ and $\rm \log [NII]\lambda6585/ H\alpha$, and gas-phase metallicity ($\rm 12 + \log(O/H)$ from \citet{Bian2018} calibrations.

\renewcommand{\arraystretch}{1.5} 
\begin{table*}[]
    \centering
    \begin{tabular}{c|c|c|c|c|c|c}
        \hline
        \hline
        Galaxy ID & $z$ & $\log M_*/M_\odot$ & $\rm \log SFR[M_\odot yr^{-1}]$ & $\rm \log OIII5008/H\beta$ & $\rm \log NII6585/H\alpha$ & $\rm 12 + \log(O/H)$ \\
        \hline
        3001 & 3.245 & $10.16 ^{+ 0.02 }_{- 0.03}$ & $ 1.09 ^{+ 0.40 }_{- 0.34 }$ & $ 0.05 ^{+ 0.17 }_{- 0.17 }$ & $ -0.24 ^{+ 0.05 }_{- 0.05 }$ & $8.72 ^{+ 0.03 }_{- 0.02 }$ \\
        1135 & 3.245 & $9.36 ^{+ 0.29 }_{- 0.32 }$ & $ 1.52 ^{+ 0.04 }_{- 0.04 }$ & $ 0.77 ^{+ 0.02 }_{- 0.02 }$ & $ -1.23 ^{+ 0.04 }_{- 0.05 }$ & $8.18 ^{+ 0.02 }_{- 0.02 }$ \\
        3174 & 3.247 & $9.02 ^{+ 0.25 }_{- 0.28 }$ & $ -0.08 ^{+ 0.18 }_{- 0.17 }$ & $ 0.48 ^{+ 0.06 }_{- 0.05 }$ & $ -0.61 ^{+ 0.09 }_{- 0.10 }$ & $8.51 ^{+ 0.03 }_{- 0.04 }$ \\
        3172 & 3.251 & $9.24 ^{+ 0.25 }_{- 0.30 }$ & $ 1.20 ^{+ 0.09}_{- 0.09 }$ & - & $ -1.11 ^{+ 0.06 }_{- 0.08 }$ & $8.27 ^{+ 0.03 }_{- 0.04 }$ \\
        3201 & 3.253 & $9.72 ^{+ 0.27 }_{- 0.39 }$ & $ 1.47 ^{+ 0.10 }_{- 0.10 }$ & $ 0.23 ^{+ 0.04 }_{- 0.04 }$ & $ -0.69 ^{+ 0.04 }_{- 0.04 }$ & $8.56 ^{+ 0.02 }_{- 0.02 }$ \\
        6003 & 3.251 & $9.11 ^{+ 0.25 }_{- 0.29 }$ & $ 0.94 ^{+ 0.07 }_{- 0.07 }$ & $ 0.64 ^{+ 0.02 }_{- 0.02 }$ & $ -1.45 ^{+ 0.11 }_{- 0.15 }$ & $8.26 ^{+ 0.02 }_{- 0.03 }$ \\
        1165 & 3.239 & $7.73 ^{+ 0.22 }_{- 0.33 }$ & $ 0.53 ^{+ 0.74 }_{- 0.69 }$ & $ 0.86 ^{+ 0.09 }_{- 0.08 }$ & $ -1.14 ^{+ 0.19 }_{- 0.30 }$ & $8.13 ^{+ 0.07 }_{- 0.10 }$ \\
        1116 & 3.231 & $7.96 ^{+ 0.29 }_{- 0.29 }$ & $ 0.09 ^{+ 0.37 }_{- 0.37 }$ & $ 0.70 ^{+ 0.08 }_{- 0.07 }$ & $ -1.75 ^{+ 0.25 }_{- 0.42 }$ & $ 8.12 ^{+ 0.12 }_{- 0.25 }$ \\
        6005 & 3.229 & $8.43 ^{+ 0.27}_{- 0.31 }$ & $ > 0.53 ^{+ 0.10 }_{- 0.09 }$ & $ >0.39 $ & $ -1.20 ^{+ 0.23 }_{- 0.38 }$ & $8.26 ^{+ 0.13 }_{- 0.20 }$ \\
        \hline
    \end{tabular}
    \caption{Measured Properties of Individual Galaxies in the MQN01 Sample.}
    \label{prop_table}
\end{table*}
\renewcommand{\arraystretch}{1} 

\section{Integrated spectra and images of MQN01 sample}
In this section we show integrated 1D spectra and composite false-color images for the MQN01 sample in Fig. \ref{spec_image_full}, complementary to those shown in Fig. \ref{spec_image}.
\begin{figure*}
  \centering
  % 自定义两行的目标高度（按需调整）
  \setlength{\rowoneht}{0.22\textheight}
  \setlength{\rowtwoht}{0.22\textheight}
  % ==== 第一行：单图，跨全宽 ====
  \includegraphics[width=0.96\linewidth]{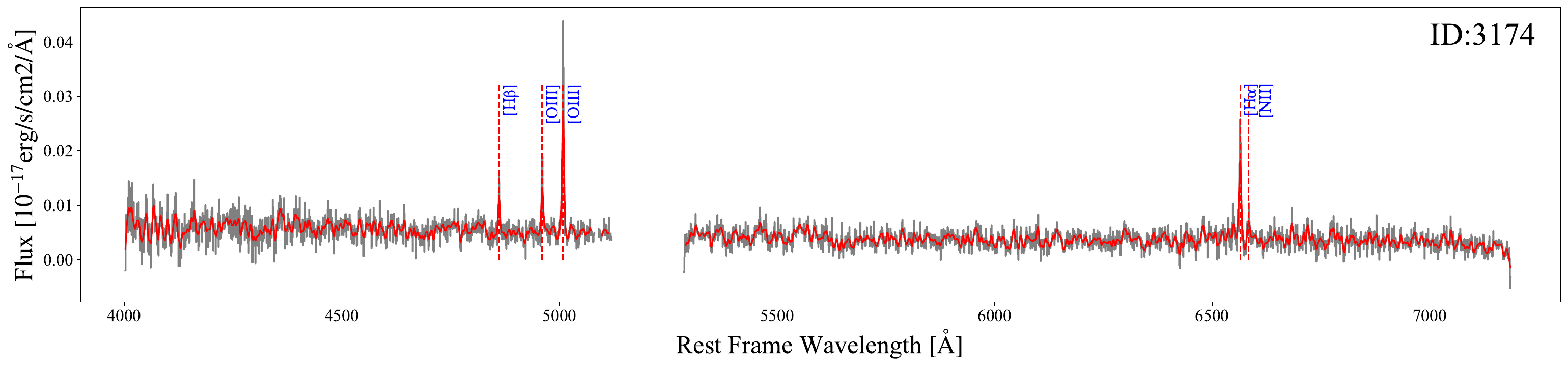}
  % ==== 第二行：两图并排、等高、零间隙 ====
  % 用两个 minipage，行末加 % 去掉空白，且不插入 \hfill
  \noindent
  \vspace{10pt}
  \begin{minipage}{0.81\linewidth}
    \includegraphics[height=\rowtwoht, width=\linewidth, keepaspectratio]{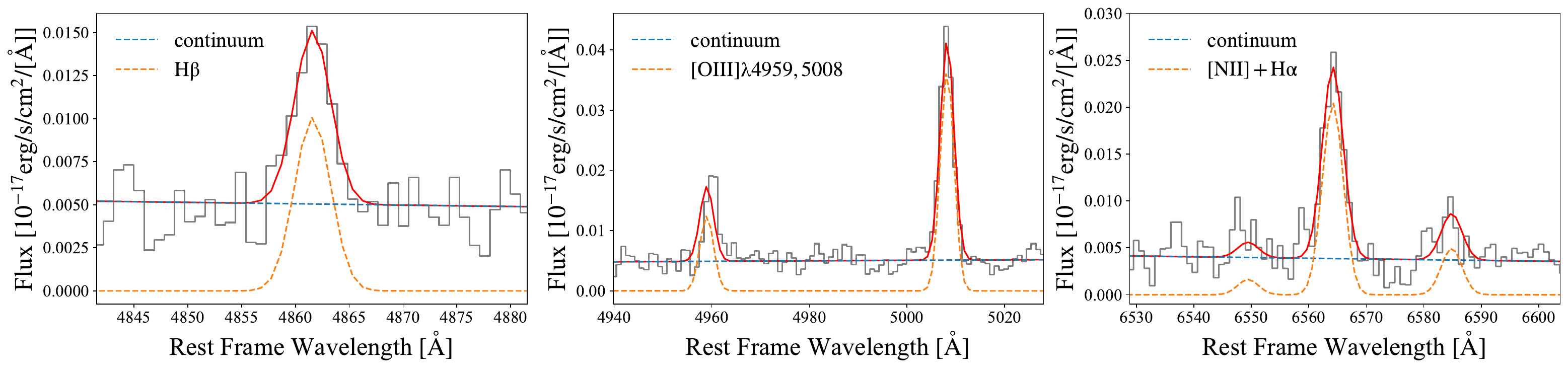}
  \end{minipage}%
  \begin{minipage}{0.19\linewidth}
    \includegraphics[height=\rowtwoht, width=\linewidth, keepaspectratio]{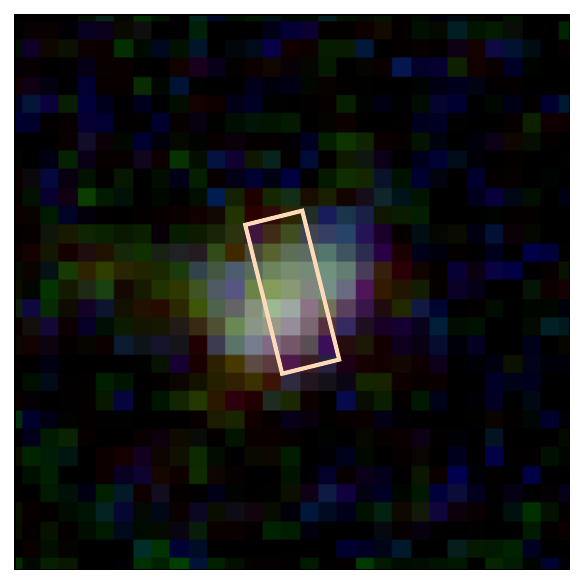}
  \end{minipage}
  % ==== 第一行：单图，跨全宽 ====
  \includegraphics[width=\linewidth]{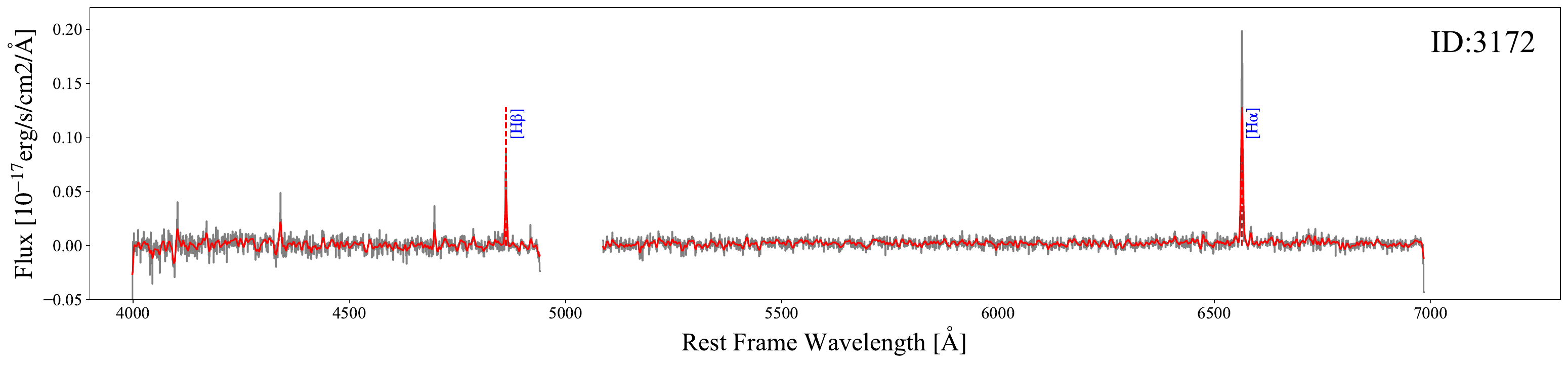}
  % ==== 第二行：两图并排、等高、零间隙 ====
  % 用两个 minipage，行末加 % 去掉空白，且不插入 \hfill
  \noindent
   \vspace{10pt}
  \begin{minipage}{0.81\linewidth}
    \includegraphics[height=\rowtwoht, width=\linewidth, keepaspectratio]{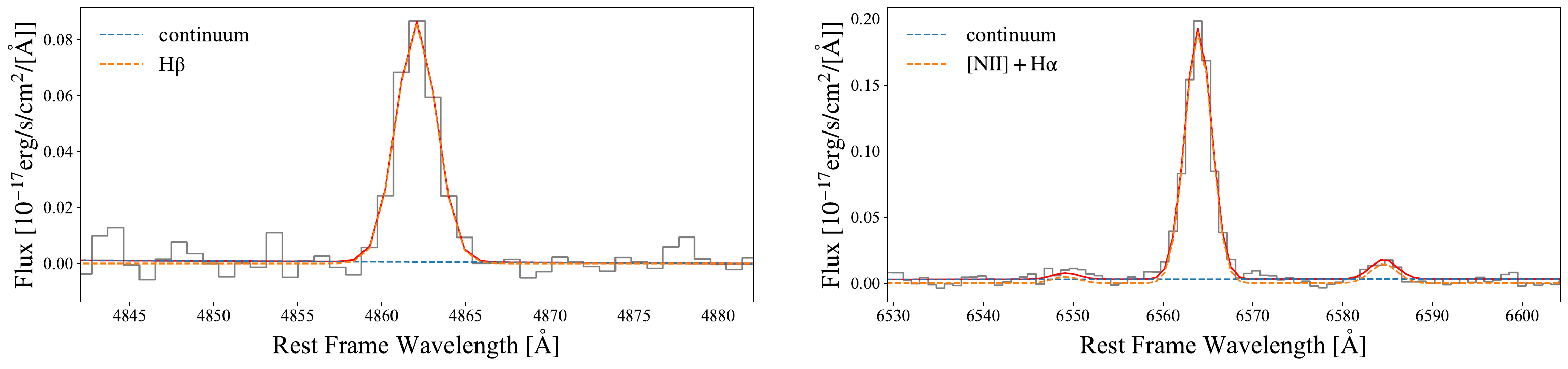}
  \end{minipage}%
  \begin{minipage}{0.19\linewidth}
    \includegraphics[height=\rowtwoht, width=\linewidth, keepaspectratio]{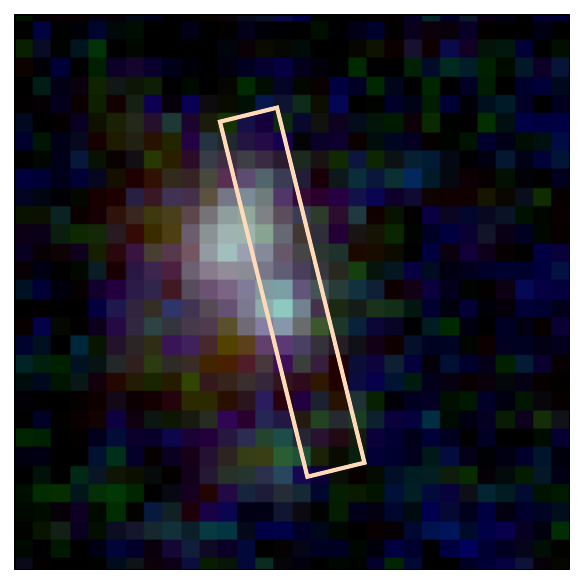}
  \end{minipage}
  \includegraphics[width=\linewidth]{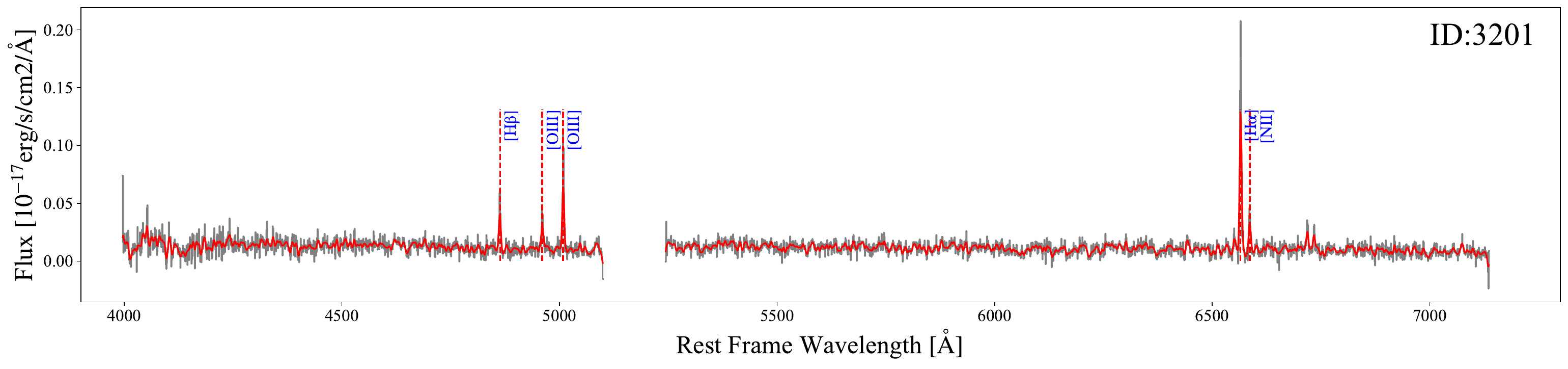}
  \noindent
  \begin{minipage}{0.81\linewidth}
    \includegraphics[height=\rowtwoht, width=\linewidth, keepaspectratio]{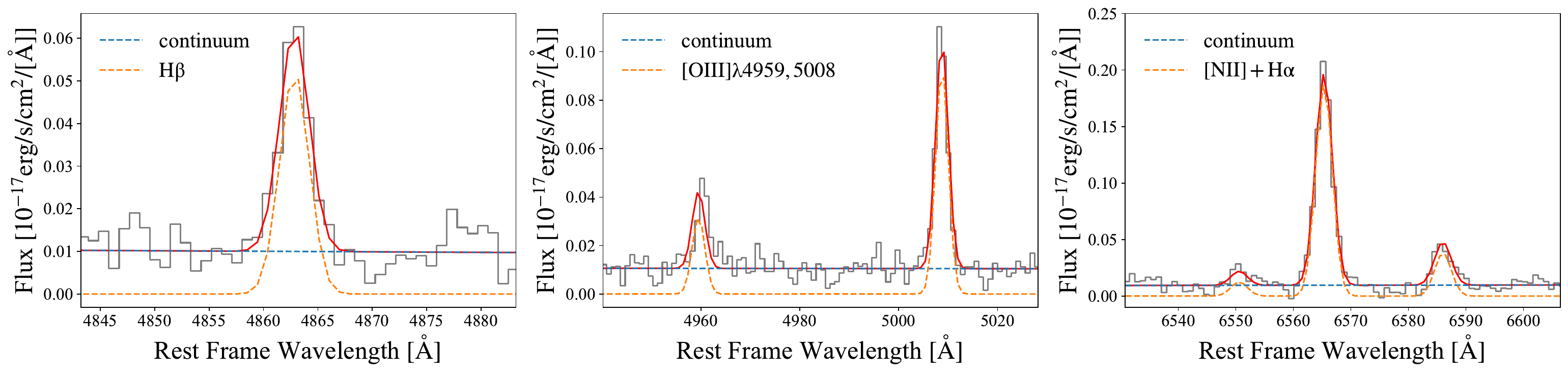}
  \end{minipage}%
  \begin{minipage}{0.19\linewidth}
    \includegraphics[height=\rowtwoht, width=\linewidth, keepaspectratio]{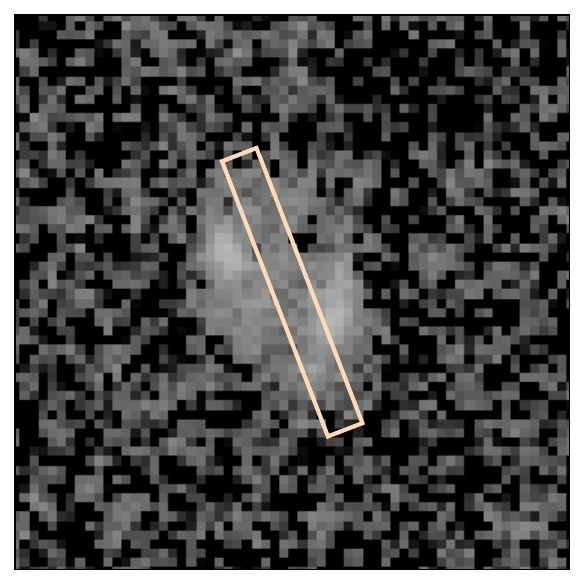}
  \end{minipage}
  \caption{Same as Fig. \ref{spec_image}, for the remaining galaxies in the sample. The galaxy with ID 3201 is not observed by JWST/NIRCam and is shown in HST F814W.}
  \label{spec_image_full}
\end{figure*}

\begin{figure*}\ContinuedFloat
  \centering
  % 自定义两行的目标高度（按需调整）
  \setlength{\rowoneht}{0.22\textheight}
  \setlength{\rowtwoht}{0.22\textheight}
  % ==== 第一行：单图，跨全宽 ====
  \includegraphics[width=\linewidth]{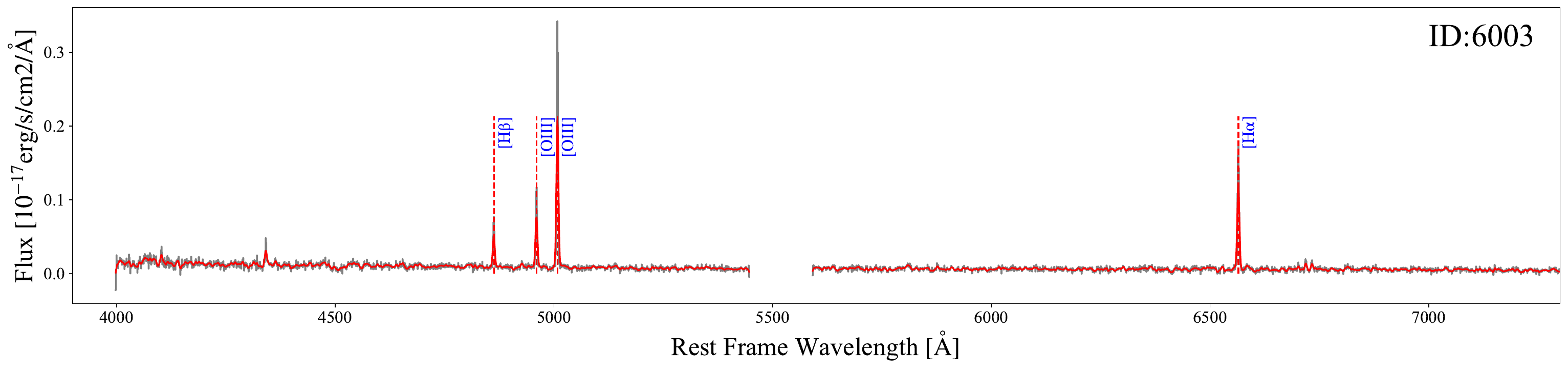}
  % ==== 第二行：两图并排、等高、零间隙 ====
  % 用两个 minipage，行末加 % 去掉空白，且不插入 \hfill
  \begin{minipage}{0.81\linewidth}
    \includegraphics[height=\rowtwoht, width=\linewidth, keepaspectratio]{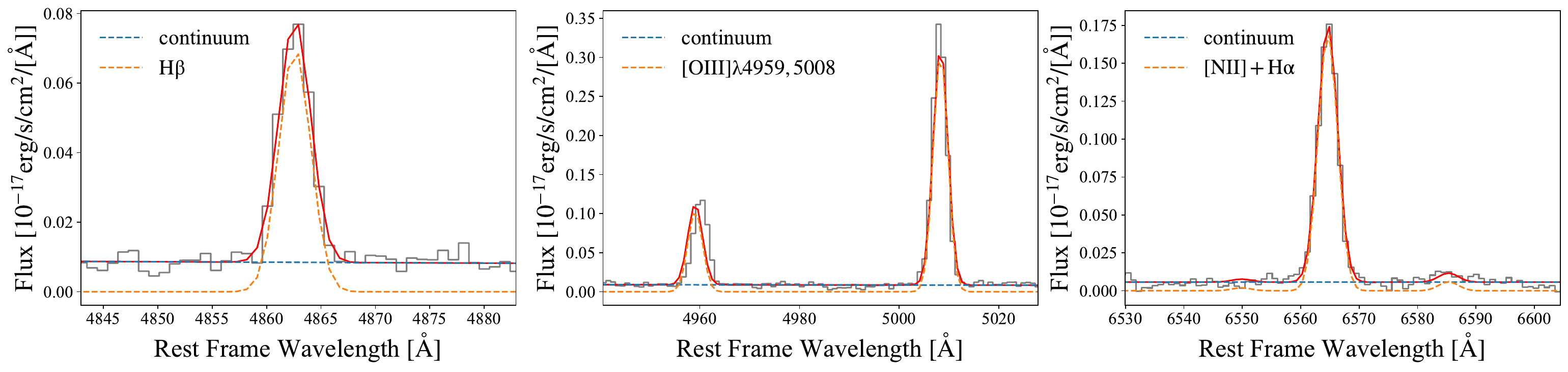}
  \end{minipage}%
  \begin{minipage}{0.19\linewidth}
    \includegraphics[height=\rowtwoht, width=\linewidth, keepaspectratio]{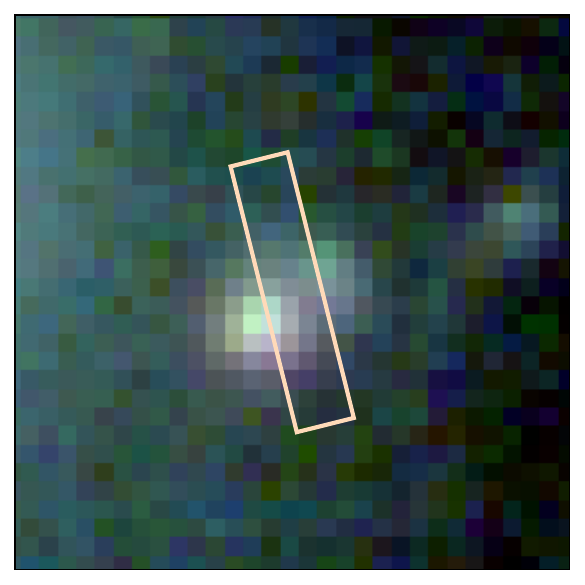}
  \end{minipage}
  % ==== 第一行：单图，跨全宽 ====
  \includegraphics[width=\linewidth]{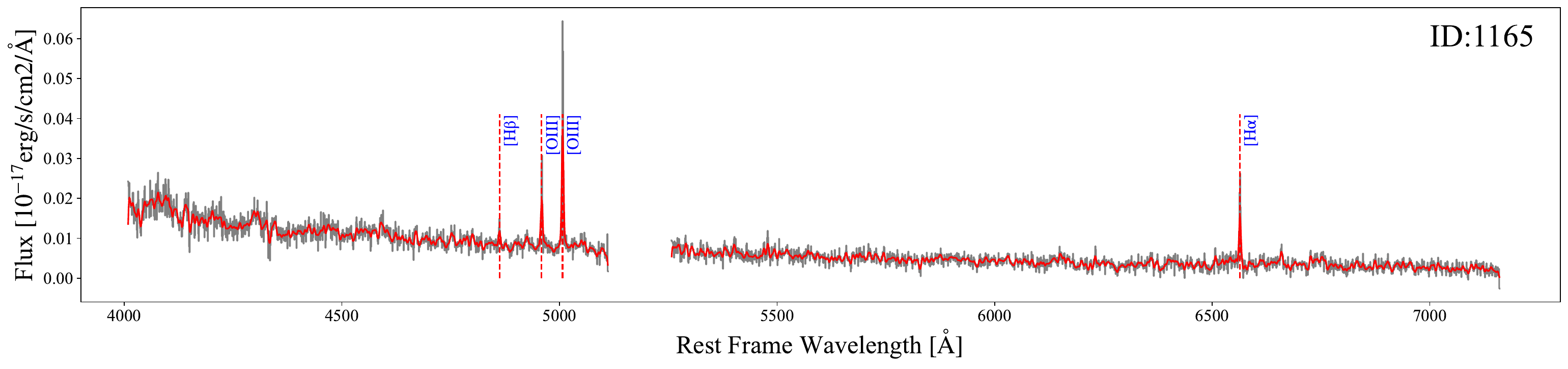}
  % ==== 第二行：两图并排、等高、零间隙 ====
  % 用两个 minipage，行末加 % 去掉空白，且不插入 \hfill
  \noindent
  \begin{minipage}{0.81\linewidth}
    \includegraphics[height=\rowtwoht, width=\linewidth, keepaspectratio]{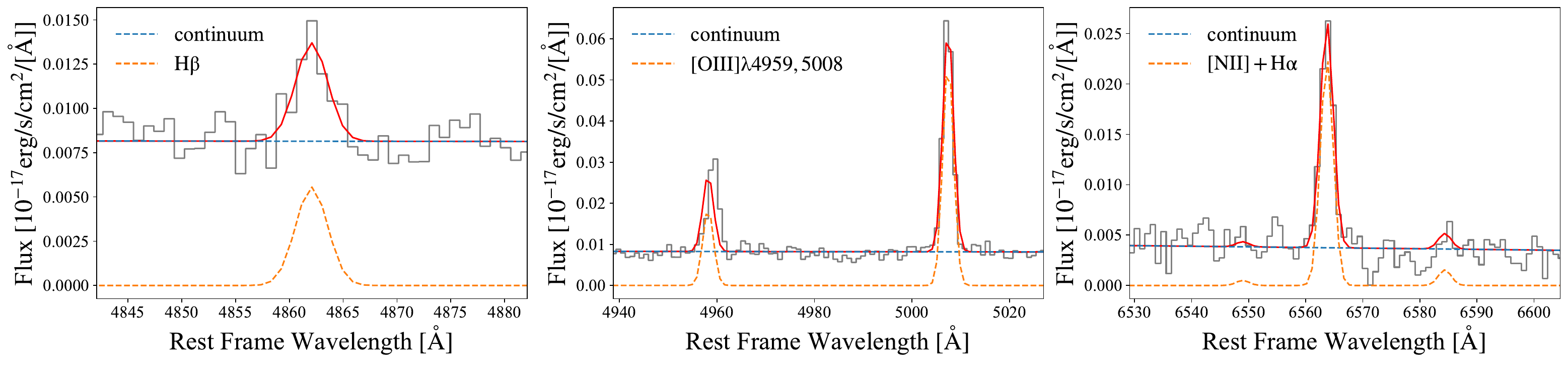}
  \end{minipage}%
  \begin{minipage}{0.19\linewidth}
    \includegraphics[height=\rowtwoht, width=\linewidth, keepaspectratio]{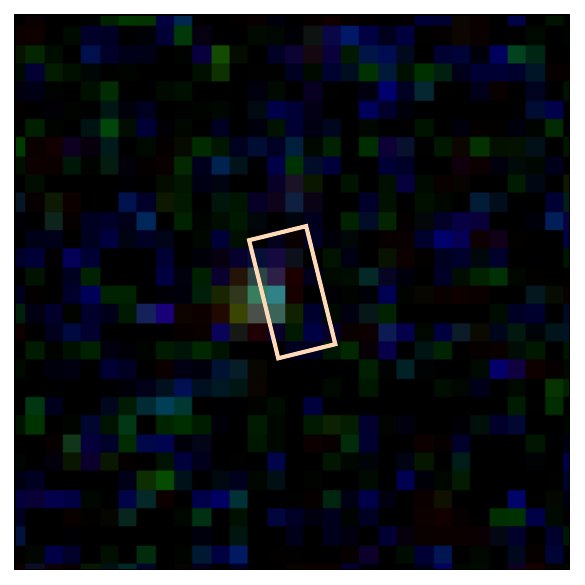}
  \end{minipage}
  \includegraphics[width=\linewidth]{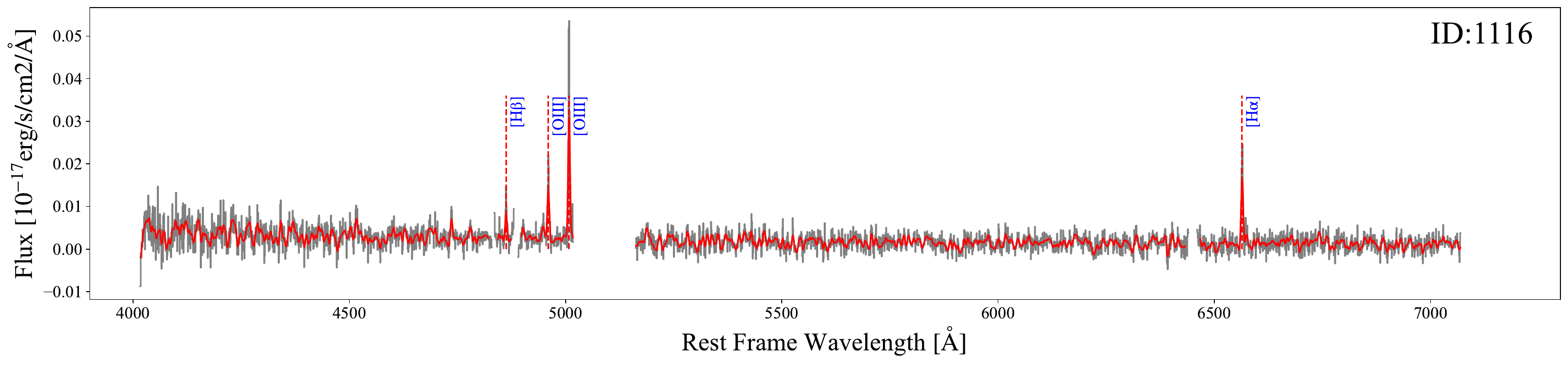}
  \noindent
  \begin{minipage}{0.81\linewidth}
    \includegraphics[height=\rowtwoht, width=\linewidth, keepaspectratio]{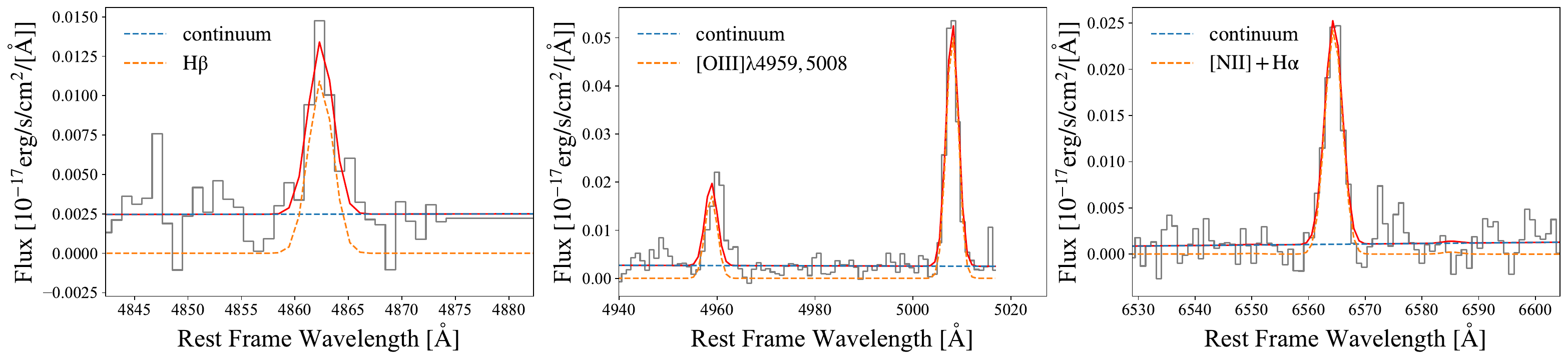}
  \end{minipage}%
  \begin{minipage}{0.19\linewidth}
    \includegraphics[height=\rowtwoht, width=\linewidth, keepaspectratio]{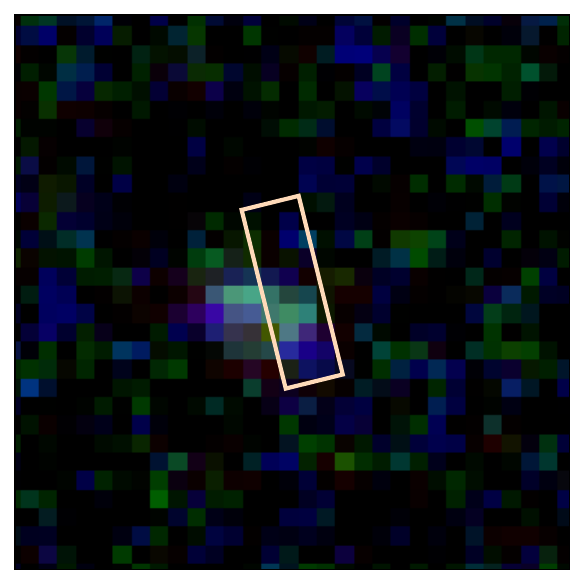}
  \end{minipage}
  \caption[Continued]{continued.}
\end{figure*}

\begin{figure*}\ContinuedFloat
  \centering
  % \newlength{\rowoneht}\setlength{\rowoneht}{0.22\textheight}
  % \newlength{\rowtwoht}\setlength{\rowtwoht}{0.22\textheight}
  \setlength{\rowoneht}{0.22\textheight}
  \setlength{\rowtwoht}{0.22\textheight}
  % 自定义两行的目标高度（按需调整）
  % ==== 第一行：单图，跨全宽 ====
  \includegraphics[width=\linewidth]{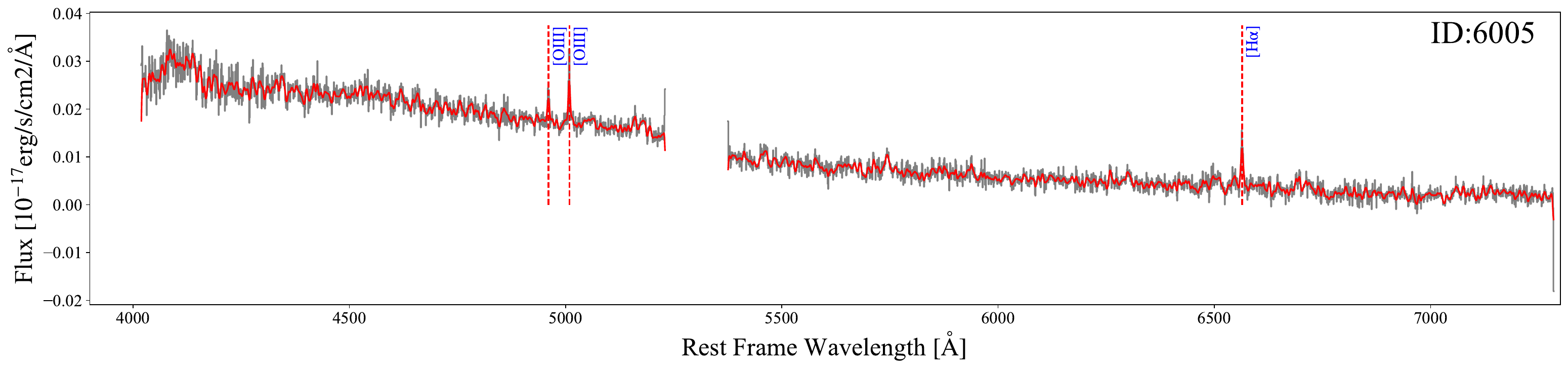}
  % ==== 第二行：两图并排、等高、零间隙 ====
  % 用两个 minipage，行末加 % 去掉空白，且不插入 \hfill
  \begin{minipage}{0.81\linewidth}
    \includegraphics[height=\rowtwoht, width=\linewidth, keepaspectratio]{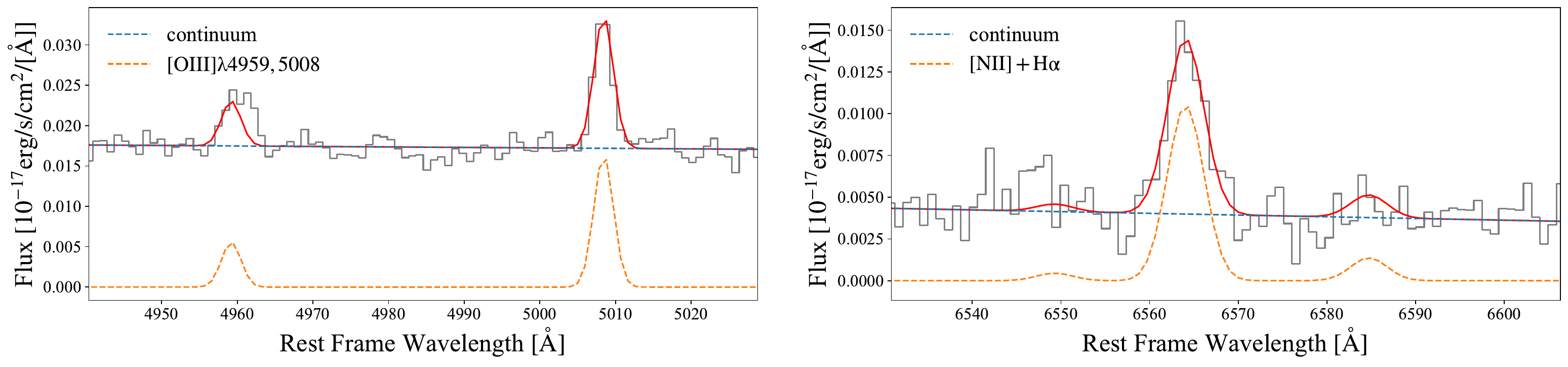}
  \end{minipage}%
  \begin{minipage}{0.19\linewidth}
    \includegraphics[height=\rowtwoht, width=\linewidth, keepaspectratio]{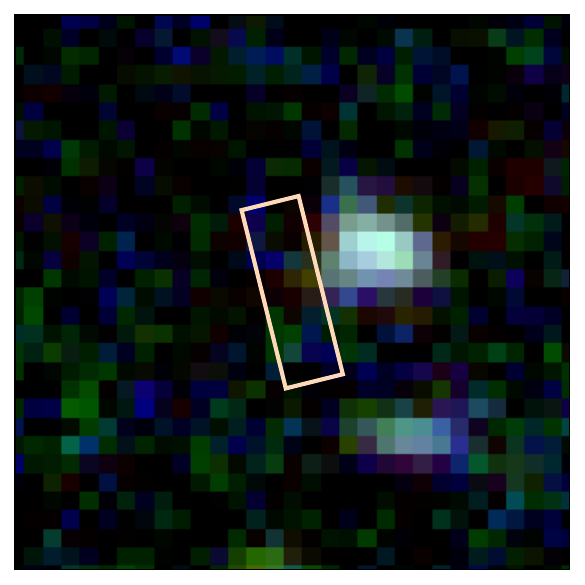}
  \end{minipage}
  \caption[Continued]{continued.}
\end{figure*}

\section{SFMS, MZR, and FZR Using Alternative Stellar Mass and SFR Estimates}
\label{new_figures_SMC}
\begin{figure*} 
    \centering
    \begin{subfigure}{0.48\textwidth}
        \includegraphics[width = \columnwidth]{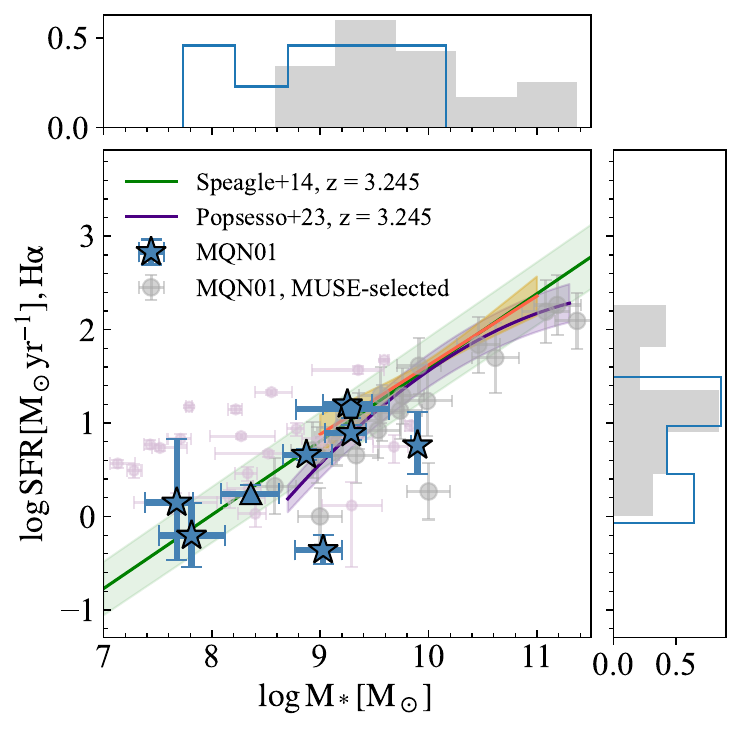}
    \end{subfigure}
    \begin{subfigure}{0.5\textwidth}
        \includegraphics[width = \columnwidth]{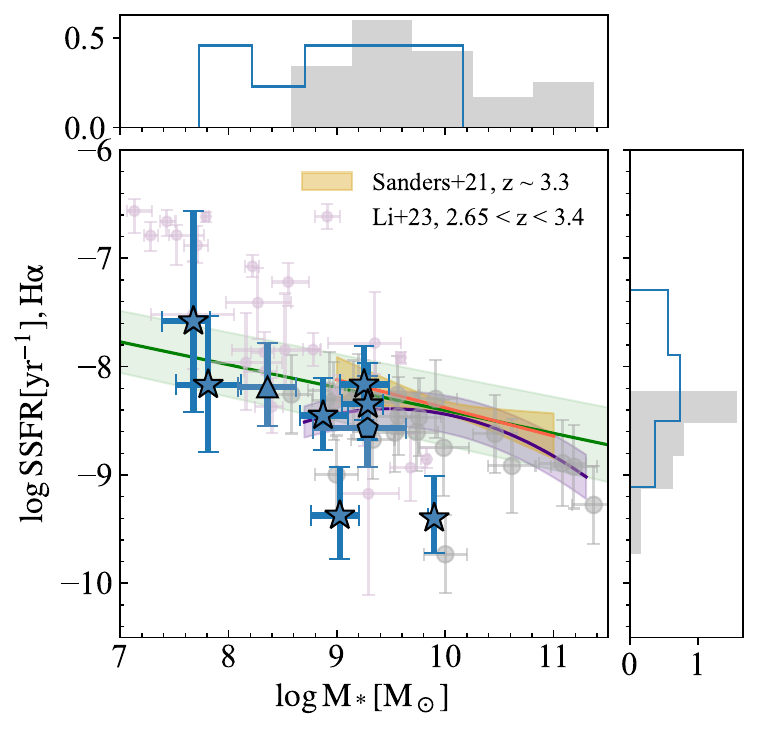}
    \end{subfigure}
    \caption{Same as Fig. \ref{SFMS}, but the stellar mass is derived from SED fitting with $Z_* = 0.004$ and SMC extinction, and SFR is calculated from $\rm H\alpha$ luminosity with the conversion factor of $10^{-41.64}.$}
    %5 galaxies reside on the SFMS, while 4 galaxies are about 0.5 - 1 dex off the SFMS.}
    \label{SFMS_SMC}
\end{figure*}

\begin{figure*} 
    \centering
    \begin{subfigure}{0.5\textwidth}
        \includegraphics[width = \columnwidth]{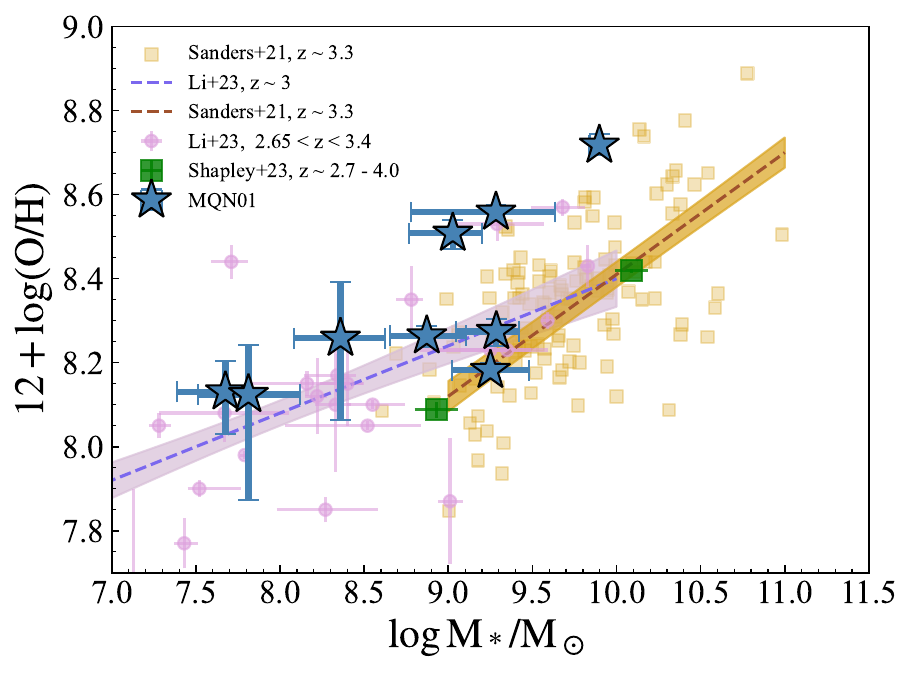}
    \end{subfigure}
    \begin{subfigure}{0.485\textwidth}
        \includegraphics[width = \columnwidth]{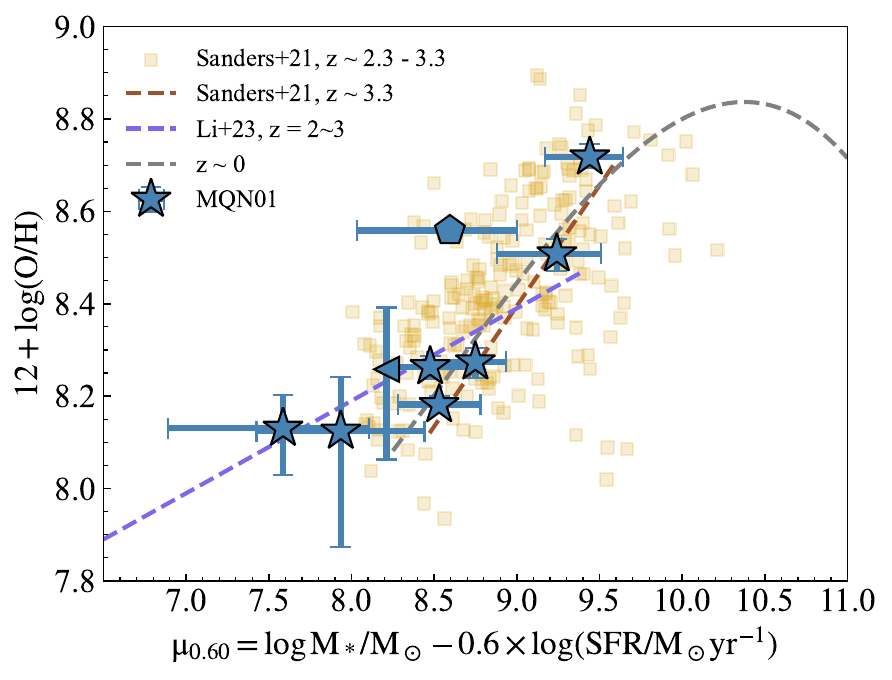}
    \end{subfigure}
    \caption{Same as Figs. \ref{MZR} and \ref{FMR}, but the stellar mass is derived from SED fitting with $Z_* = 0.004$ and SMC extinction, and SFR is calculated from $\rm H\alpha$ luminosity with the conversion factor of $10^{-41.64}.$}
    %5 galaxies reside on the SFMS, while 4 galaxies are about 0.5 - 1 dex off the SFMS.}
    \label{MZR_FMR_SMC}
\end{figure*}

In this section, we reproduce the star formation main sequence (SFMS), mass-metallicity relation (MZR), and fundamental metallicity relation (FZR) with stellar mass derived from $Z_* = 0.004$ + SMC SED fitting and SFR with conversion factor of $10^{-41.64}$. The relations are shown in Figs. \ref{SFMS_SMC} and \ref{MZR_FMR_SMC}.
The overall trends and our main conclusions, metal enhancement compared with field galaxies and less difference on the FZR,  remain unchanged.

\end{appendix}

\end{document}